\documentclass[10pt,journal,compsoc]{IEEEtran}

\usepackage{cite}
\usepackage{amsmath,amssymb,amsfonts,amsthm}
\usepackage{tabularx,booktabs}
\usepackage{subfigure}
\usepackage{graphicx}
\usepackage{textcomp}
\usepackage{url}
\usepackage{diagbox}
\usepackage{makecell}
\usepackage{hyperref}

\usepackage{xcolor}
\usepackage{booktabs}
\usepackage[ruled,linesnumbered]{algorithm2e}
\usepackage{multirow}
\usepackage{enumitem}
\usepackage{todonotes}
\usepackage{color, soul}
\usepackage[most]{tcolorbox}
\usepackage{cleveref}

\newcommand{\model}{\text{KPER}\xspace}
\newcommand{\modelv}{\text{KPER}}

\SetKwComment{Comment}{$\triangleright$\ }{}
\definecolor{block1}{rgb}{0.2, 0.2, 0.2}
\definecolor{content}{RGB}{245, 245, 245}
\definecolor{purple}{RGB}{201, 187, 207}

\crefname{section}{§}{§§}
\Crefname{section}{§}{§§}

\def\emb{\boldsymbol} 
\DeclareMathOperator\mlp{mlp}
\DeclareMathOperator\mean{mean}

\ifCLASSINFOpdf
\else
\fi

\begin{document}

\title{\huge Knowledge-aware Neural Networks with Personalized Feature Referencing for Cold-start Recommendation}


\author{
    Xinni Zhang, 
    Yankai Chen, 
    Cuiyun Gao, 
    Qing Liao, 
    Shenglin Zhao, 
    and Irwin King,~\IEEEmembership{Fellow,~IEEE}
    
    \IEEEcompsocitemizethanks{
    \IEEEcompsocthanksitem X. Zhang, C. Gao (Corresponding Author), Q. Liao, and S Zhao are with the Department of Computer Science, Harbin Institute of Technology, Shenzhen, China. Y. Chen and I. King are with the Department of Computer Science, Chinese University of Hong Kong, Hong Kong SAR, China.\protect\\
    E-mail: \{zhangxinni.hit, zsl.zju\}@gmail.com, \\ 
    \{ykchen, king\}@cse.cuhk.edu.hk,  \{gaocuiyun, liaoqing\}@hit.edu.cn.
    }
}

%
%

\markboth{Journal of \LaTeX\ Class Files,~Vol.~14, No.~8, August~2015}%
{Shell \MakeLowercase{\textit{et al.}}: Bare Demo of IEEEtran.cls for Computer Society Journals}
%

\IEEEtitleabstractindextext{%

\begin{abstract}

Incorporating knowledge graphs (KGs) as side information in recommendation has recently attracted considerable attention. Despite the success in general recommendation scenarios, prior methods may fall short of performance satisfaction for the cold-start problem in which users are associated with very limited interactive information. Since the conventional methods rely on exploring the interaction topology, they may however fail to capture sufficient information in cold-start scenarios. To mitigate the problem, we propose a novel Knowledge-aware Neural Networks with Personalized Feature Referencing Mechanism, namely KPER. Different from most prior methods which simply enrich the targets' semantics from KGs, e.g., product attributes, KPER utilizes the KGs as a ``semantic bridge'' to extract feature references for cold-start users or items. Specifically, given cold-start targets, KPER first probes semantically relevant but not necessarily structurally close users or items as adaptive seeds for referencing features. Then a Gated Information Aggregation module is introduced to learn the combinatorial latent features for cold-start users and items. 
Our extensive experiments over four real-world datasets show that, KPER consistently outperforms all competing methods in cold-start scenarios, whilst maintaining superiority in general scenarios without compromising overall performance, e.g., by achieving 0.81\%$\sim$16.08\% and 1.01\%$\sim$14.49\% performance improvement across all datasets in Top-10 recommendation.

\end{abstract}

}

\maketitle

\IEEEdisplaynontitleabstractindextext
\IEEEpeerreviewmaketitle


\IEEEraisesectionheading{\section{\textbf{Introduction}}
\label{introduction}}
\IEEEPARstart{W}{ith} the explosive growth of web information, recommender systems (RSs), as a useful tool for information filtering~\cite{ying2018graph, 9101889}, nowadays are versatile to various Internet databases and applications.
Among recent advances in online recommender models, incorporating \textit{Knowledge Graphs} (KGs) as side information has lately attracted considerable attention~\cite{2018_RippleNet,2019_KGAT,2020_ckan, chen2022modeling}.
Instead of simply leaning on user-item interaction data, with the rich and heterogeneous relational information in KGs to supplement additional semantics, KG-based RS methods thus have the potential to provide more precise, diverse, and explainable information organization and recommendation.

Despite their progress in general recommendation scenarios, existing KG-based models still fall short of performance satisfaction for the \textit{cold-start problem}.
Generally, cold-start problem describes the scenario where users tend to have \textit{extremely limited} interactive transactions~\cite{2021_KGPL,2022_metakg, 2020_metahin}.
This problem is ubiquitous in applications where user and item databases are frequently updated~\cite{2021_KGPL}, e.g., E-commerce platforms and social networks.
Although KGs are capable of contributing auxiliary semantic enrichment, e.g., product attributes and profiles, the learning process however can hardly gather sufficient \textit{interactive information}~\cite{wiki} in the cold-start scenario, restricting the performance of KG-based recommender methods.

 \begin{figure}[t]
  \centering
  \includegraphics[scale=0.45]{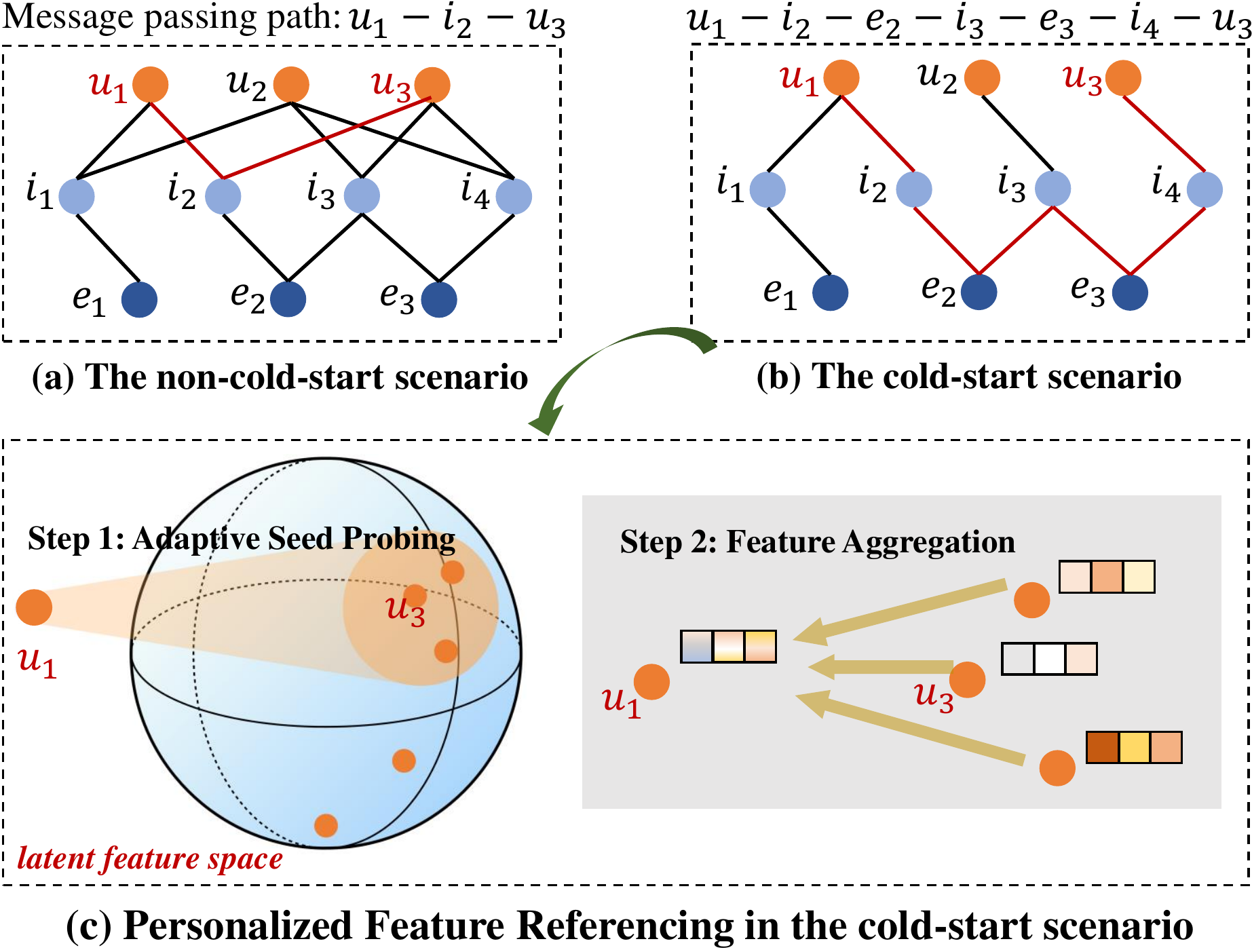}
    \caption{Motivation illustration (best view in color).}
  \label{fig:motivation}
  \vspace{-0.15cm}
\end{figure}

Technically, \textit{prior models dedicate to designing better knowledge extraction paradigms, which however are constrained by topology sparsity for information propagation}.
Specifically, early works~\cite{2018_MCRec, cao2019unifying, 2019_RuleRec} follow certain manually defined patterns (i.e., meta-paths), to summarize information from KGs with specific semantic meanings.
Other models~\cite{2018_DKN, 2018_KSR,2019_KGAT} design additional regularization loss terms to jointly learn user-item latent features and KG embeddings for both sides of interaction data and KGs.
However, in a cold-start scenario with sparse user-item interactions, these methods may collapse to merely optimize KG embeddings and fail to learn sufficient collaborative signals~\cite{2020_ckan} for modeling the interactive information of target users and items.
Recent works~\cite{2019_KGCN, 2019_KGNN_LS,chen2022learning} adopt Graph Neural Networks (GNNs) to conduct explicit message passing from KGs to eventually enrich user-item embeddings. 
One similar inadequacy is that current GNN-based models heavily rely on topology exploration to propagate distant knowledge.
As shown in Fig.~\ref{fig:motivation}(a), if the user-item interactions are sufficient, the interactive information, a.k.a., \textit{collaborative signals}~\cite{he2020lightgcn,2020_ckan,2021_AttentiveKG, yang2022hrcf, yang2022hicf}, can easily propagate back and forth, e.g., via the path ``$u_1$$-$$i_2$$-$$u_3$'', while the KG information is ``icing on the cake'' to further boost the semantics incidentally. 
But in sparse interaction structures shown in Fig.~\ref{fig:motivation}(b), such an information propagation mechanism thus hardly produces prompt and effective information dissemination via the long passing path ``$u_1$$-$$i_2$$-$$e_2$$-$$i_3$$-$$e_3$$-$$i_4$$-$$u_3$''.

To address these issues, we propose \textit{\underline{K}nowledge-aware Neural Networks with \underline{P}ersonalized F\underline{E}ature \underline{R}eferencing} (\model).
\model specializes to alleviate the cold-start problem without compromising overall recommendation performance. 
We propose a novel and effective mechanism named \textbf{Personalized Feature Referencing}.
 As shown in Fig.~\ref{fig:motivation}(c), this mechanism utilizes the KG as a ``semantic bridge'' so as to reference representative latent features to users and items that originally suffer from insufficient learning of interactive information. 
Orthogonal to the aforementioned methods, our model can get rid of the interaction sparsity constraint, and achieve the embedding enrichment in the \textit{latent semantic space}.
Technically, \model establishes the sequential pipeline with the following three modules where:
\begin{enumerate}[leftmargin=*]
    \item \textbf{Interactive Information Encoding.}
    We first employ it to summarize the latent features from the observed interactions to sketch the representations for both users and items.
    
    \item \textbf{Attentive Knowledge Encoding.}
    We further develop multi hops of knowledge extraction from the KG whilst maintaining knowledge diversity across different users and items.
    
    \item \textbf{Personalized Feature Referencing.}
        Integrating both sides of the information learned so far, for a given user (or item), we then employ \textbf{Adaptive Seed Probing} to explicitly detect user (or item) representatives.
    Intuitively, these representatives possess sufficient interaction information and share \textit{semantically similar} features with the given user (or item), but may be \textit{structurally distant}.
    We provide an effective \textbf{Continuous Approximation} approach to introduce the more stable model training, followed by the \textbf{Gated Information Aggregation}.
    Equipped with such augmented feature referencing, the cold-start problem can thus be substantially alleviated. 
\end{enumerate}

\noindent{\textbf{Empirical Results.}}
We conduct the comprehensive experiments and  analyses on four real-world datasets:
\begin{itemize}[leftmargin=*]
    \item In general recommendation scenarios, \model outperforms the state-of-the-art Collaborative-filtering-based and KG-based recommender models by 0.81\%$\sim$16.08\% and 1.01\%$\sim$14.49\% performance improvement across all datasets, in terms of \textit{Recall} and \textit{Precision} metrics of Top-10 recommendation task. 
    
    
    \item In cold-start recommendation scenarios, our model \model consistently presents the performance superiority over all competing models, across various user or item groups with different levels of interaction sparsity.
    
    \item Our detailed empirical analysis and ablation study not only demonstrates the effectiveness of our proposed Personalized Feature Referencing Mechanism, but also validates the solidity of each model component in contributing to holistic model performance.
\end{itemize}

\noindent{\textbf{Discussion.}}
It is worthwhile mentioning that there are few KG-based methods targeted at alleviating the cold-start problem via \textit{Meta-Learning}~\cite{2020_metahin, 2022_metakg} and \textit{Graph Semi-Supervised Learning}~\cite{2021_KGPL}.
In this work, we focus on proposing an effective neural network architecture, i.e., \model. 
Hence, we believe that \model is \textit{motivationally similar} but \textit{technically orthogonal} to these learning paradigms. 
We provide their model introduction in~\cref{related_work} and also include them for performance comparison in experiments.

\noindent{\textbf{Organization.}}
The rest of this paper is organized as follows. 
We first review the related work in~\cref{related_work}.
Then we present the detailed explanation of our model \model in~\cref{sec:method} and attach model analysis in~\cref{sec:analysis}.
We report the extensive experiments in~\cref{exp} with conclusion in~\cref{conclusion}.


\section{\textbf{Related Work}}
\label{related_work}

\subsection{\textbf{Cold-start Recommendation}}
\noindent\textbf{Asking to Rate.}
Early solutions to address the cold-start issue mainly focus on the \textit{asking to rate}.
This is a explicitly-interactive operation that profiles target ``cold'' users by presenting items to them~\cite{nadimi2014cold}.
They can be generally categorized into (1) adaptive and (2) non-adaptive methods. 
oncretely, non-adaptive methods present the identical item lists to all users for the explicit ratings, which seeks to improve the model performance globally.
To achieve this, various strategies have been deployed, such as Entropy~\cite{rashid2008learning}, Random, Popularity, and Balanced strategies~\cite{rashid2002getting}.
As for the adaptive methods, the item list asking for ratings is consistent with each target user's opinions and preference.
Therefore, the rated items with personalized orderings enable the cold-start recommendation accuracy to be more effectively improved; but the interview process will also be more complicated with increased costs.
Related methods are referred in~\cite{rashid2002getting,aimeur2006better,rashid2008learning}.

\noindent\textbf{Incorporating with Content Information.}
Apart from solely relying on the rating information, several models are proposed to combine preference and additional content information~\cite{agarwal2009regression,wang2011collaborative,wang2015collaborative}.
However, the content part of these methods is typically generative~\cite{gopalan2014content,wang2011collaborative,wang2015collaborative} such that the models are optimized to ``explain'' the content rather than maximizing the recommendation accuracy~\cite{volkovs2017dropoutnet}.
One recent work~\cite{volkovs2017dropoutnet} starts to employ deep neural networks to leverage preference and content information without explicitly relying on both being present.
Another recent work~\cite{2019_CCCC} proposes a deep attentive neural architecture to model both content and interactive information for generalization ability enhancement.
However, one major issue is that the \textit{intra-relationships} of the additional information are not captured.
To tackle this, KGs, as they are capable of modeling various kinds of relational knowledge and side information within the graph structures, enable KG-based recommender models to receive a wide range of attention recently.

\subsection{\textbf{Knowledge Graph-based Recommendation}}
Studying KGs as a kind of side-information has aroused interests in various applications including recommender systems. Existing KG-based recommender models are mainly in three types: (1) path-based~\cite{2014_Hete-MF, 2017_FMG, 2018_MCRec, 2019_RuleRec}, (2) embedding-based~\cite{2016_CKE, 2018_DKN, 2018_KSR}, and (3) hybrid methods~\cite{2018_RippleNet, 2019_KGAT, 2021_AttentiveKG, 2020_ckan, chen2022modeling}.

\noindent\textbf{Path-based Methods.} 
Path-based methods explore various connecting patterns among items in KGs (i.e., meta-paths or meta-graphs) to provide additional guidance for recommendations. 
Generally, the generation of such connecting patterns heavily relies on either the path generation algorithm~\cite{2018_HERec} or manual creation~\cite{2018_MCRec}. 
Although path-based methods naturally bring interpretability and explanation into the recommendation process, it can be difficult to design such patterns with limited domain knowledge.
Thus for the large-scaled and complicated KGs, it is impractical to perform the exhaustive path retrieval and generation, while the selected paths make a great impact on the final recommendation performance.

\noindent\textbf{Embedding-based Methods.}
These methods adopt knowledge graph embedding (KGE) algorithms~\cite{2017_KGE_survey} to directly exploit the semantic information in KGs, and then enrich the representations of users and items. 
For instance, DKN~\cite{2018_DKN} utilizes TransD~\cite{2015_TransD} to jointly process the KGs and learn item embeddings. 
KV-MN~\cite{2018_KSR} proposes a knowledge-enhanced key-value memory network which adapts TransE~\cite{2013_TransE} to depict user preference in sequential recommendation. 
KTUP~\cite{cao2019unifying} adopts TranH~\cite{wang2014knowledge} to learn the KG embeddings via hyperplanes to further improve the recommendation performance.


\noindent\textbf{Hybrid Methods.} 
They combine the above two techniques to achieve the state-of-the-art performance, and thus have attracted much attention in recent years. 
They usually apply iterative information propagation under graph neural network framework to generate the entity representations for information enrichment. 
For example, CKAN~\cite{2020_ckan} employs a heterogeneous propagation strategy along multi-hop links to encode knowledge associations for users and items. 
CG-KGR~\cite{2021_AttentiveKG} customizes KG information with the proposed collaborative guidance mechanism which encapsulates historical interaction information to provide personalized recommendation.
However, all of these methods are not designed for cold-start recommendation.
Two models MetaHIN and MetaKG~\cite{2020_metahin, 2022_metakg} exploit the power of \textit{meta-learning} paradigm. 
Technically, they model the preference learning for each user as a single meta-learning task. 
However, one major issue is that optimizing every single meta-learner for each user from the interaction record would incur expensive computational costs.
Another latest model KGPL~\cite{2021_KGPL} adopts \textit{graph semi-supervised learning} to develop \textit{pseudo-labeling} via random walk. 
This essentially is a graph simulation to increase interaction density by assigning pseudo-positive items to users.
However, it starts with the structure exploration and thus may be influenced by the initial state of interaction sparsity, while those \textit{de facto} negative labeling computed from the sparse substructures would thus perturb and destabilize the model training for recommendation.
We include them in the experiments for performance comparison.



\section{\textbf{Problem Formulation}}
\label{sec:pre}
User-item interactions can be represented by a bipartite graph ${\mathcal{G}_1=\{(u,r',i)|u\in\mathcal{U},i\in\mathcal{I}}\}$, where $r'$ generalizes all user behaviours, e,g., \textit{click, purchase}, as one relation type between user $u$ and item $i$. 
${\mathcal{U}}$ and ${\mathcal{I}}$ are the sets of users and items, respectively.
We use $y_{ui}=1$ to indicate the user $u$ has interacted with the item $i$, otherwise $y_{ui}=0$. 
We have the knowledge graph ${\mathcal{G}_2=\{(h,r,t)|h,t\in\mathcal{E},r\in\mathcal{R}\}}$ in which $\mathcal{E}$ and $\mathcal{R}$ are the collections of entities and relations.
Each knowledge triple $(h,r,t)$ denotes that there is a relation $r$ connecting from head entity $h$ to tail entity $t$. 
Similar to~\cite{2019_KGAT,2020_ckan}, each item can be matched with an entity in the knowledge graph (i.e., $\mathcal{I}\subseteq\mathcal{E}$). For ease of interpretation, we unify the user-item bipartite graph $\mathcal{G}_{1}$ and knowledge graph $\mathcal{G}_{2}$ into the \textit{collaborative knowledge graph} (CKG)  $\mathcal{G}=(\mathcal{E}^\prime, \mathcal{R}^\prime)$, where $\mathcal{E}^\prime=\mathcal{E} \cup \mathcal{U}$ and $\mathcal{R}^\prime=\mathcal{R}\cup\{r'\}$. 

\textbf{Notations.} We use bold lowercase, bold uppercase, and calligraphy letters for vectors, matrices, and sets, respectively. 
Non-bold ones are used to denote graph nodes or scalars.
We list the key notations
in Table \ref{tab:Symbol}.

\textbf{Task Formulation.}
Given the CKG $\mathcal{G}$, we aim to train a recommender model $\mathcal{F}(\cdot, \cdot|\Theta,\mathcal{G})$, with the capability of estimating  probability $\hat{y}_{ui}$ that user $u$ will engage with item $i$, i.e., $\hat{y}_{ui}=\mathcal{F}(u,i|\Theta,\mathcal{G})$, where $\Theta$ is the parameter set in $\mathcal{F}$.

\begin{table}[tbp]
\caption{{Symbols and Description.}}
  \label{tab:Symbol}
  \begin{center}
  \begin{tabular}{c|m{6.0cm}}
    \hline
    \textbf{Notation} & \textbf{Description} \\
    \hline
    \hline
    $\mathcal{U, I}$ & The sets of users, items. \\
    \hline
    $\mathcal{E, R}$ & The sets of entities, relations. \\
    \hline
    $r'$ & Generalized user behaviour type. \\
    \hline
    $(h,r,t)$ & The knowledge triple. \\
    \hline
    $\mathcal{G}_1, \mathcal{G}_2$ & The user-item bipartite graph and item-side knowledge graph. \\
    \hline
    $\mathcal{G}=(\mathcal{E}^\prime, \mathcal{R}^\prime)$ & The collaborative knowledge graph contains $\mathcal{G}_1$ and $\mathcal{G}_2$, with $\mathcal{E}^\prime=\mathcal{E}\cup\mathcal{U}$,  $\mathcal{R}^\prime=\mathcal{R}\cup\{r'\}$. \\
    \hline
    $o$ & The uniform placeholder for user symbol $u$ and item symbol $i$. \\
    \hline
    $\emb{{v}}_{o} \in \mathbb{R}^d$ & The embedding of $o$.\\
    \hline
    $\emb{{v}}_{o}^{k} \in \mathbb{R}^d$ & The knowledge-aware embedding of $o$ at hop $k$.\\
    \hline
    $\emb{{v}}_h, \emb{{v}}_t, \emb{{v}}_r \in \mathbb{R}^d$ & The embedding of head entity $h$, tail entity $t$, and relation $r$.\\
    \hline
    $\mathcal{E}_o^{k}$ & The $k$-hop entity-neighbor set of $o$. \\
    \hline
    $\mathcal{T}_o^{k}$ & The sampled $k$-hop triple-neighbor set of $o$. \\
    \hline
    $l$ & The neighbor sample size in interactive information encoding. \\
    \hline
    $K$ & The depth of knowledge extraction. \\
    \hline
    $\mathcal{S}$ & The initial seed pool. \\
    \hline
    $\{\emb{{t}}_{x}\}_{x=1}^{|\mathcal{S}|}$ & The embedding table. \\
    \hline
    $\emb{t}_o^{*}\in \mathbb{R}^{2d}$ & The feature referencing embedding of $o$. \\
    \hline
    $\emb{z}_{o} \in \{0,1\}^{|\mathcal{S}|}$ & The indicator for seed selection of $o$. \\
    \hline
    $\emb{v}_o^{+}\in \mathbb{R}^{2d}$ & The gated aggregation embedding of $o$. \\
    \hline

  \end{tabular}
  \end{center}
\end{table}

\section{\textbf{\model Methodology}}
\label{sec:method}

 \begin{figure*}[t]
  \centering
  \includegraphics[scale=0.54]{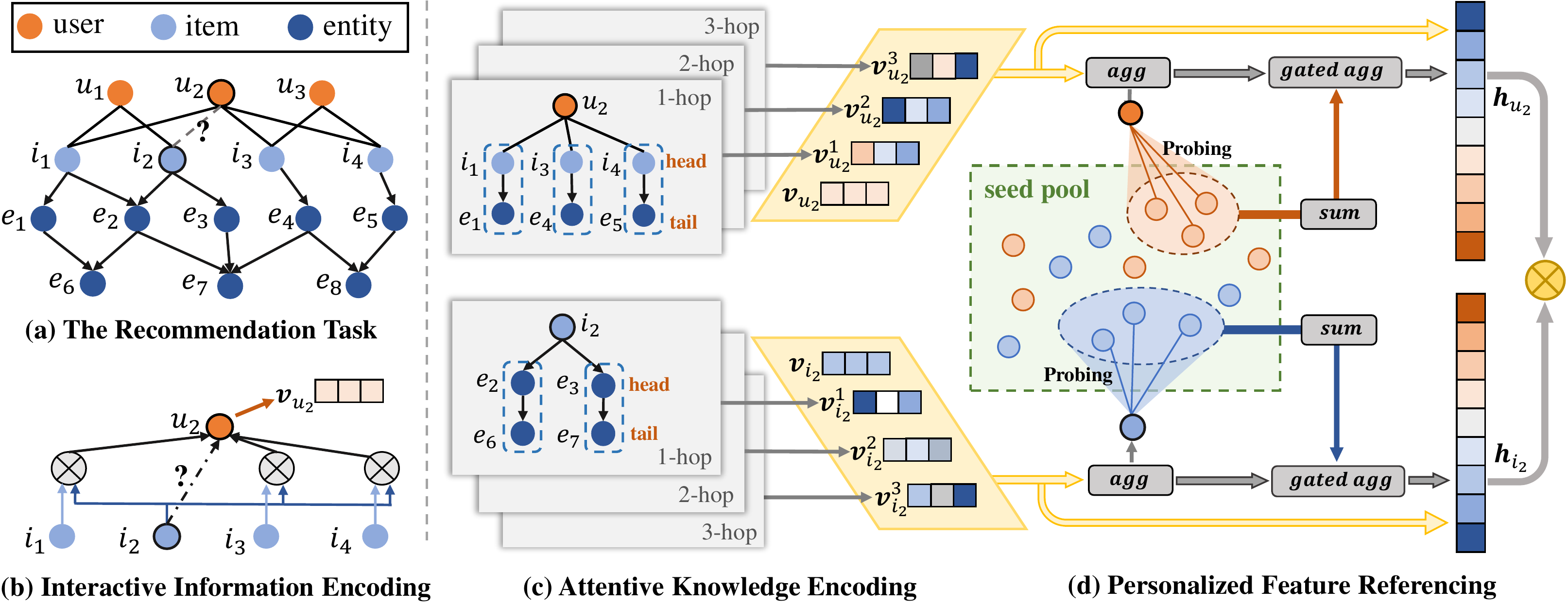}
  \caption{\model Illustration: predicting the interaction probability between user $u_2$ and item $i_2$ (best view in color).}
  \label{fig:model}
\end{figure*}


\subsection{\textbf{Model Overview}}
We detail the proposed \model model framework, with the architecture illustration in Fig. \ref{fig:model}. \model consists of three main components in which: 
(1) \textbf{Interactive Information Encoding} adaptively embeds the interactive information from observed historical interactions (\cref{sec:User_Embedding});
(2) \textbf{Attentive Knowledge Encoding} employs an attentive neural module to generate knowledge-aware embeddings for both users and items (\cref{sec:Knowledge-aware_Attentive_Embedding}); 
(3) \textbf{Personalized Feature Referencing} then combines the information from modules (1) and (2) to firstly probe representative user-item candidates, and then use the corresponding embeddings for feature referencing to alleviate the cold-start recommendation.

\subsection{\textbf{Interactive Information Encoding}}
\label{sec:User_Embedding}
In cold-start scenarios, it is essential to learn interactive information from the observed neighbors, as this can directly reflect the users' preference towards items and items' targeting costumers in users.
Following this intuition, we firstly introduce our \textit{Interactive Information Encoding}.
Let $o$ denote a node placeholder for users (or items), concretely, the neighboring interactive information of $o$ is defined as:
\begin{equation} 
\label{eq:his1}
  \emb{{v}}_o =\sum_{o'\in\mathcal{N}(o)} \alpha(\emb{{v}}_{o_{p}}, \emb{{v}}_{o'}) \emb{{v}}_{o'},
\end{equation}
where $\emb{{v}}_{o}, \emb{{v}}_{o_{p}}, \emb{{v}}_{o'} \in \mathbb{R}^d$ denote the embeddings of $o$, $o$'s target node $o_{p}$, and the historical interacted counterpart $o'$ from $o$'s neighbor set $\mathcal{N}(o)$, e.g., node $u_2$, node $i_2$ and node set $\{i_1, i_3, i_4\}$ in Fig.~\ref{fig:model}(b), respectively.
Coefficient $\alpha(\emb{{v}}_{o_{p}}, \emb{{v}}_{o'})$ is calculated as:
\begin{equation}
\label{eq:his2}
  \alpha(\emb{{v}}_{o_{p}}, \emb{{v}}_{o'}) =\frac{\exp(\emb{{v}}_{o_{p}}^{\top} {\emb{{v}}_{o'}})} {\sum_{o^{''}\in\mathcal{N}(o)} \exp(\emb{{v}}_{o_{p}}^{\top} {\emb{{v}}_{o''}})}.
\end{equation}
Intuitively, the encoding process of node $o$'s embedding dynamically adjusts the information compositions from historical interactions, i.e., $\mathcal{N}(o)$, whilst considering their matching relevance with the specific node $o'$, i.e., $\alpha(\emb{{v}}_o, \emb{{v}}_{o'})$.

\subsection{\textbf{Attentive Knowledge Encoding}}
\label{sec:Knowledge-aware_Attentive_Embedding}

In addition to learning interaction information, we also explore the KG structure and explicitly propagate the knowledge therein to enrich the embedding semantics for both users and items.
Specifically, we propose \textit{Attentive Knowledge Encoding}, which employs the (1) \textit{K-hop Knowledge Sampling} and (2) \textit{Knowledge Encoding} accordingly.

\subsubsection{K-hop Knowledge Sampling}
We first sample node $o$'s distant entity-neighbors, via \textit{recursively} exploring the knowledge triplets, e.g., $(h,r,t) \in \mathcal{G}$, in $K$ hops of KG substructures:
\begin{equation}
    \mathcal{E}_o^k = \{t|(h,r,t)\in\mathcal{G} \text{ and } h\in\mathcal{E}_o^{k-1} \}, \quad k=1,2,\ldots,K,
\end{equation}
where $k$ indicates the distance from the initial neighbor set of users and items, e.g., $\mathcal{E}_u^{0}$ and $\mathcal{E}_i^{0}$, that are defined as follows:
\begin{equation}
\mathcal{E}_u^{0} =\{e|(i,e)\in\mathcal{G}, (u,i)\in\mathcal{G}\}, 
\mathcal{E}_i^{0}  =\{e|(u,e)\in\mathcal{G}, (u,i)\in\mathcal{G}\}.
\end{equation}

For example, as shown in Fig.2 (a), let $o$ = $u_1$, the initial entity-neighbor set and 1-hop entity-neighbor set of $u_1$ is $\mathcal{E}_{u_1}^{0}=\{i_1, i_2\}$ and $\mathcal{E}_{u_1}^{1}=\{e_1, e_2, e_3\}$, respectively. 
Then based on these KG substructures, we define the fixed-size knowledge triple sampling set from $\mathcal{E}_{o}^{k-1}$ as follows:
\begin{equation}
    \mathcal{T}_o^k = \{(h,r,t)| h\in \mathcal{E}_o^{k-1}\}, \quad k=1,2,\ldots,K.
\end{equation}

\subsubsection{Knowledge Encoding}
Based on sampled knowledge associations, e.g., $(h,r,t) \in \mathcal{T}_o^k$, we learn the knowledge-aware embeddings for $o$: 
\begin{equation}
    \label{eq:kgex2}
    \emb{{v}}_o^k = \sum_{(h,r,t) \in \mathcal{T}_o^k} \pi(\emb{{v}}_{h}, \emb{{v}}_{r}) \emb{{v}}_{t}, \quad k=1,2,\ldots,K,
\end{equation}
where coefficient $\pi(\emb{{v}}_{h}, \emb{{v}}_{r})$ is attentively calculated as:
\begin{equation}
\label{eq:weight1}
    \pi(\emb{{v}}_{h}, \emb{{v}}_{r}) = \frac{\exp(\mlp(\emb{{v}}_{h}||\emb{{v}}_{r}))}{\sum_{(h^{\prime},r^{\prime},t^{\prime}) \in \mathcal{T}_o^k} \exp(\mlp(\emb{{v}}_{h'}||\emb{{v}}_{r'}))}.
\end{equation}
Notation $||$ denotes the concatenation operation.
$\emb{{v}}_h, \emb{{v}}_t \in \mathbb{R}^d$ are the embeddings of head entity $h$ and tail entity $t$. $\emb{{v}}_r \in \mathbb{R}^d$ is the embeddings of relation $r$.
$\mlp(\cdot)$ is implemented by a two-layer MLP with the nonlinear activation RELU \cite{2011_relu}. 

Notice that the attention mechanism in encoding our knowledge-aware embeddings (Equations~(\ref{eq:kgex2})-(\ref{eq:weight1})) provides the \textit{relevance calibration} functionality to tail entity $t$ by considering the semantics of head entity $h$ and relation $r$.
For example, knowledge triplets (\textit{SpaceX}, \textit{Founded by}, \textit{Elon Musk}) and (\textit{Iron Man 2}, \textit{Guest Performed by}, \textit{Elon Musk}) exhibit two different roles of tail entity ``\textit{Elon Musk}'', which thus calibrates its knowledge relevance in embedding encoding. 
Furthermore, by increasing $k$ from 1 to $K$, we can achieve the high-order knowledge encoding in deeper KG structures, which substantially enriches the embedding semantics to improve the cold-start recommendation performance.

After learning the embeddings from interactive (\cref{sec:User_Embedding}) and KG sides (\cref{sec:Knowledge-aware_Attentive_Embedding}), we then condense them into a $2d$-dimensional embeddings as follows:
\begin{equation} 
\label{eq:encoding1}
  \emb{{v}}^*_o = \emb{{v}}_o || \mean(\{ \emb{{v}} _o^k\}_{k=1}^{K}).
\end{equation}
Intuitively, embedding $\emb{{v}}_o^* \in \mathbb{R}^{2d}$ collects information from both data sides, making the preparation for next-stage \textit{Personalized Feature Referencing}, which specializes the customized feature enrichment to nodes with sparse interactions.




\subsection{\textbf{Personalized Feature Referencing}}
\label{sec:Dynamic_shared_Embedding}

\noindent\textbf{Motivation.} 
The conventional methodology to alleviate the cold-start issue usually focuses on the simulation on interaction graph structures to increase interaction density, e.g., via random walk~\cite{2021_KGPL}.
However, it relies on the local topology exploration, and may thus fail to find those \textit{structurally distant} but \textit{semantically similar} node candidates.
On the contrary, thanks to the rich semantic supplementation from both interactive and KG sides, we seek to achieve such a \textbf{candidate probing} process mainly in the embedding space, where both structural and semantic information are well encoded.


\noindent\textbf{Challenges.}
A straightforward implementation would be performing certain clustering algorithms, e.g, K-means \cite{1967_kmeans}, to locate node candidates that are near the cluster centroids, as these centroids are usually more representative to present unique semantics than other ordinary nodes.
Despite the straightforwardness, we argue that this trivial implementation may contain several inadequacies, which may lead to sub-optimal recommendation performance:
(1) These clustering methods usually proceed independently of the personalized recommendation. Since users have diverse item-preferences (likewise, items possess various attracting-groups to users), a unified candidate selection approach may however be too coarser-grained especially for personalized recommendation.
(2) Moreover, aside from the difference in users' preferences and items' attractivity, the number of the desired candidates should also vary for different users and items.
However, the existing clustering methods may hardly provide such customization, e.g., by detecting whether a user has a wide or narrow interest coverage;
otherwise, the manual pruning may be extremely resource-expensive.

To address these issues, we propose an adaptive candidate probing method, which
enables the effective searching for \textit{semantic focused} and \textit{coverage pruned} \textbf{seeds} (\textit{a.k.a.} users or items that are used for information referencing).
To furnish a tractable model training, we then introduce an advanced \textit{continuous approximation} approach for it and finally we explain the information aggregation procedure to enrich the user-item representations for cold-start recommendation.

\subsubsection{Adaptive Seed Probing}
To avoid the exhaustive search over the whole node corpus, we firstly collect a small range of nodes as the initial \textit{seed pool} $\mathcal{S}$\footnote{Please notice that in practice we implement $\mathcal{S}$ into two disjoint sets for users and items, respectively. Without loss of generality, their corresponding seed probing processes can be interpreted via the placeholder $o$ as follows. }, where $|\mathcal{S}| \ll |\mathcal{G}|$.
Since cold-start interaction graphs usually follow the long-tailed distribution \textit{w.r.t.} node degrees~\cite{park2008long}, similar to work~\cite{wu2021self}, we obtain $\mathcal{S}$ by 
excluding tailed-nodes and pick those with high popularity (implementation details are reported in~\cref{sec:cold_start_exp} of experiments).
Then for all seeds in $\mathcal{S}$, we build an additional embedding table as follows: 
\begin{equation}
\{{\emb{t}}_{1}, {\emb{t}}_{2}, \cdots, {\emb{t}}_{{|\mathcal{S}|}}\}.
\end{equation}
These embeddings demonstrate the representative features of highly-interacted users and items, providing the functionality of information aggregation for ``cold'' nodes with very limited interactions. 
Similar to~\cite{weston2014memory, 2020_prehash}, we separate these \textit{feature referencing} embeddings out of those learned in previous encoding sections, e.g., knowledge encoding in Equation~(\ref{eq:kgex2}), mainly to prevent issues such as \textit{over-training}.

Then for any node $o$, we expect our model to train an indicator $\emb{z}_{o} \in \{0,1\}^{|\mathcal{S}|}$, in which each 1-valued entry indicates that the seed in this position is selected.
Furthermore, to prevent seed \textit{redundancy} or even \textit{noise}, these seeds are not only representative but also expected to be \textit{concise} and \textit{precise}.
To achieve this, we can minimize the objective in Equation~(\ref{eq:binary}) to sparsify $\emb{z}_o$ (i.e., as few 1-valued entries as possible), by penalizing the number of non-zero entries in $\emb{z}_o$:
\begin{equation}
\label{eq:binary}
||\emb{z}_o||_0 =  \sum_{x=1}^{|S|}\mathbb{I}[(\emb{z}_o)_x \neq 0].
\end{equation}
Here $\mathbb{I}$ counts the number of 1-valued entries in $\emb{z}_o$.
Then in the holistic training objective, we can further maximize the recommendation accuracy, which eventually makes a trade-off between recommendation performance and seed-searching pruning.
However, directly optimizing the objective in Equation~(\ref{eq:binary}) is computationally intractable,  mainly because its non-differentiability.
Thus to address this issue and provide more stable model training, we introduce an advanced solution with the \textit{continuous approximation} in the following section.

\subsubsection{Continuous Approximation}
To provide the accordant continuous approximation, we first learn a sequence of continuous scores $\emb{\beta}_o$, measuring the selection confidence of seeds in $\mathcal{S}$:
\begin{equation}
    \label{eq:weight2}
\begin{aligned}
    \emb{\beta}_o =\exp(\emb{W} \emb{v}_o^* + \emb{b}),
\end{aligned}
\end{equation}
Where $ \emb{W} \in \mathbb{R}^{|\mathcal{S}|\times 2d}$ and $ \emb{b} \in \mathbb{R}^{|\mathcal{S}|}$ are the learnable parameters.
$\emb{v}_o^*$ is learned from previous stage in Equation~(\ref{eq:encoding1}), bridging the information learnt from historic interactions (\cref{sec:User_Embedding}) and knowledge associations (\cref{sec:Knowledge-aware_Attentive_Embedding}).

Then, based on the learned $\emb{\beta}_o$, one of trustworthy approaches for relaxing discrete parameters~\cite{2017_l0sparse, 2021_learn_to_drop} is to explicitly smooth each binary entity in $\emb{z}_{o}$, correlated by certain random variables, e.g., $\emb{\gamma}_o$ $\in$ $\mathcal{R}^{|S|}$.
Specifically, $\emb{\gamma}_o$ is a concrete random variable drawn from:
\begin{equation}
\label{eq:concrete}
\begin{aligned}
 \emb{\gamma}_o = \sigma((\log\emb{\xi} - \log\emb{(\emb{1}-\emb{\xi})} + \log\emb{\beta}_o)/\tau), 
  \emb{\xi} \sim \text{\small{Uniform}}(\emb{0},\emb{1}),
\end{aligned}    
\end{equation}
where $\sigma(\cdot)$ is the sigmoid function and $\tau$ is the temperature. 
Since the entries of $\emb{\gamma}_o$ ranges in $(0,1)$, to force those entries of undesired seeds to be exactly 0, we further rescale their ranges to $(\eta,1)$, with the parameter $\eta < 0$:
\begin{equation}
\label{eq:rscaling}
    \Bar{\emb{\gamma}}_o = (1 - \eta) \emb{\gamma}_o + \eta, 
\end{equation}
which enables the negative entries to be rounded to 0. Then we obtain our approximated indicator $\Bar{\emb{z}}_o$ as follows:
\begin{equation}
\label{eq:cut}
    \Bar{\emb{z}}_o = \max(\Bar{\emb{\gamma}}_o, \emb{0}).
\end{equation}
By developing such continuous relaxation for computational stable model training, we can effectively learn a combinatorial seed allocation for personalized information referencing, whilst achieving the target of search space pruning in $\mathcal{S}$.

\subsubsection{Gated Information Aggregation}
Based on the learned seed indicator $\bar{\emb{z}}_o$, we now proceed to the combinatorial information aggregation for feature referencing as follows:
\begin{equation}
    \label{eq:sharing}
    \emb{t}_o^{*} = 
    \begin{cases}
    \emb{t}_a, & \text{if $o$ $\in$ $\mathcal{S}$ with id $a$},  \\
    \sum_{x=1}^{|\mathcal{S}|} \text{softmax}_x(({\bar{\emb{z}}}_o)_x) {\emb{t}}_x, & \text{otherwise}.
    \end{cases}
\end{equation}
In order to automatically aggregate information for the ``cold'' nodes with sparse interactions, inspired by the Gated Recurrent Unit (GRU)~\cite{2014_GRU},  we design the neural \textit{Gated Information Aggregation} module:
\begin{equation}
\label{eq:dy_sgg2}
    \emb{v}_o^{+} =  \frac{\exp(c_1)} {\exp(c_1) + \exp(c_2)} \emb{v}_o^{*} +  \frac{\exp(c_2)} {\exp(c_1) + \exp(c_2)} \emb{t}_o^{*},
\end{equation}
where gate signals $c_1$ and $c_2$ balances the information contribution, and they can be calculated as follows:
\begin{equation}
\label{eq:weight3}
    c_1 = \emb{q}^\top \sigma(\emb{W}_{c}\emb{v}_o^* + \emb{b}_{c}),  \text{  and  }
    c_2 = \emb{q}^\top \sigma(\emb{W}_{c}\emb{t}_o^* + \emb{b}_{c}).
\end{equation}
$\emb{W}_{c}$ $\in$ $\mathbb{R}^{d\times 2d}$, $\emb{b}_{c}, \emb{q}$ $\in$ $\mathbb{R}^d$, are the learnable parameters.
As we can observe from Equation~(\ref{eq:dy_sgg2}), our proposed fusion gate mechanism can adaptively control the information integration from two parts, i.e., $\emb{v}_o^{*}$ and $\emb{t}_o^{*}$.
And it can prevent \textit{information overload} issue especially for those nodes which are with \textit{neutral} or even \textit{far-from-urgent} cold-start recommendation demands.
As we will show later in Experiment~\cref{sec:rq1}, our model can thus perform consistently well across all user-item groups with different interaction density.

\subsection{\textbf{Model Prediction and Optimization}}
\label{sec:optimize}
\noindent\textbf{Model Prediction.}
We package the multiple embedding segments from \textit{Interactive Information Encoding} (\cref{sec:User_Embedding}), \textit{Knowledge Encoding} (\cref{sec:Knowledge-aware_Attentive_Embedding}), and \textit{Personalized Feature Referencing} (\cref{sec:Dynamic_shared_Embedding}) into a unified representation:
\begin{equation}
     \emb{h}_o = \emb{v}_o ||\{ \emb{v} _o^k\}_{k=1}^{K} || \emb{v}_o^{+}.
\end{equation}
Then to estimate the matching probability between a user $u$ and item $i$, we naturally adopt the \textit{inner-product} as follows: 
\begin{equation}
    \hat{y}_{u,i} = \emb{h}_u^\top \emb{h}_i.
\end{equation}

\

\noindent\textbf{Model Optimization.}
Our objective function for model optimization mainly consists of two loss terms: 
\begin{itemize}[leftmargin=*]
\item $\mathcal{L}_{CE}$: the cross-entropy loss reconstructs the observed interactive topology;
\item $\mathcal{L}_{SP}$: a\,penalty\,term\,sparsifies\,the\,personalized\,seed\,probing.
\end{itemize}
Concretely, we implement $\mathcal{L}_{CE}$:
\begin{equation}
\mathcal{L}_{CE} = \sum_{u\in\mathcal{U}}\big(\sum_{i\in{\{ \mathcal{Y}_u^+, \mathcal{Y}_u^- \}}} \mathcal{J}(y_{u,i}, \hat{y}_{u,i})\big).
\end{equation}
$\mathcal{Y}_u^+$ denotes the positive (ground-truth) interacted items of user $u$ and $\mathcal{Y}_u^-$ is the negative item set, where $\lvert \mathcal{Y}_u^+ \rvert = \lvert \mathcal{Y}_u^- \rvert$.
As for $\mathcal{L}_{SP}$, we first introduce its formulation in Equation~(\ref{eq:sploss}) and report the detailed derivation process in later section~\cref{sec:ana}.
\begin{equation}
\label{eq:sploss}
\mathcal{L}_{SP} = \sum_{o \in \{\mathcal{U} \cup \mathcal{I}\}} \sum_{x=1}^{|\mathcal{S}|} \sigma\big(\log\big((\emb{\beta}_o)_x\big) - \tau \text{log}(-\eta)\big).
\end{equation}
Based on these two loss terms, we build up our final objective function as follows:
\begin{equation}
\mathcal{L} = \mathcal{L}_{CE} + \lambda_1 \mathcal{L}_{SP} + \lambda_2 \Vert \Theta \Vert_2^2,
\end{equation}
where $\Theta$ is the set of all trainable parameters, and $\Vert \Theta \Vert_2^2$ is the $L_2$-regularizer parameterized by $\lambda$ to avoid over-fitting. 
So far, we have introduced all the technical parts of our model \model. 
We attach the pseudo-codes in Algorithm~\ref{alg:model} and formally report the model analysis in the following section.

\begin{algorithm}[t]
\small
\caption{\model algorithm}
\label{alg:model}
\KwIn{Collaborative knowledge graph $\mathcal{G}$ contains $\mathcal{G}_1$, $\mathcal{G}_2$;
    sampled triple-neighbors: $\{{\mathcal{T}^{}_o}^k\}_{k=1}^{K}$; \qquad \qquad
    trainable parameters $\Theta$: $\{\emb{{v}_e}\}_{e\in\mathcal{E}}$, $\{\emb{{v}_r}\}_{r\in\mathcal{R}}$, $\{\emb{{t}}_{x}\}_{x=1}^{|\mathcal{S}|}$,
    $\{\emb{W}, \emb{b}\}$ in Eq.\ref{eq:weight1}, \ref{eq:weight2}, and \ref{eq:weight3}; \qquad \qquad
    hyperparameters: $l, K, d, B, \eta, \tau, \lambda_1, \lambda_2$.}
\KwOut{Prediction function $\mathcal{F}(\cdot,\cdot|\Theta,\mathcal{G})$.}
\While{{\model} not converge}{
    \For{$(u,i)$ {in} $\mathcal{G}$}{
        $\emb{{v}}_u, \emb{{v}}_i \leftarrow$ encode interactive info\Comment*[r]{Eq.\ref{eq:his1}-\ref{eq:his2}}\
        \For{$k=1,2,\ldots, K$}{
            $\emb{{v}}_u^k, \emb{{v}}_i^k \leftarrow$ encode K-hop knowledge\Comment*[r]{Eq.\ref{eq:kgex2}-\ref{eq:weight1}}
        }
        $\emb{{v}}^*_u = \emb{{v}}_u || \mean(\{ \emb{{v}} _u^k\}_{k=1}^{K})$\; $\emb{{v}}^*_i = \emb{{v}}_i || \mean(\{ \emb{{v}} _i^k\}_{k=1}^{K})$\;
        $\emb{\beta}_u, \emb{\beta}_i \leftarrow$ learn continuous scores\Comment*[r]{Eq.\ref{eq:weight2}}
        $\Bar{\emb{z}}_u, \Bar{\emb{z}}_i \leftarrow$ approximate indicator\Comment*[r]{Eq.\ref{eq:concrete}-\ref{eq:cut}}
        $\emb{t}_u^{*},\emb{t}_i^{*} \leftarrow$  embed feature referencing\Comment*[r]{Eq.\ref{eq:sharing}}
        $\emb{v}_u^{+}, \emb{v}_i^{+} \leftarrow$ aggregate gated info\Comment*[r]{Eq.\ref{eq:dy_sgg2}-\ref{eq:weight3}}
        $\hat{y}_{u,i} \leftarrow$ generate estimated matching score\;
        $\mathcal{L} \leftarrow$ compute loss and optimize \model model\;
    }
}
\KwRet{$\mathcal{F}$}.
\end{algorithm}


\section{\textbf{Analysis of Optimizing Seed Probing}}
\label{sec:analysis}

\label{sec:ana}
The initial objective function for optimizing binary indicators can be formulated as follows:
\begin{equation}
\label{eq:inti_loss}
  \mathcal{L}_{SP} = \sum_{o \in \{\mathcal{U} \cup \mathcal{I}\}} \sum_{x=1}^{|S|}\mathbb{I}[(\emb{z}_o)_x \neq 0],
\end{equation}
which however, as we have mentioned before, it could be computationally intractable to directly optimize.
To tackle this issue, we introduce the continuous relaxation.
Then the corresponding objective function can be re-formulated:
\begin{equation}
\label{eq:re_l0}
  \mathcal{L}_{SP} = \sum_{o \in \{\mathcal{U} \cup \mathcal{I}\}} \sum_{x=1}^{|\mathcal{S}|} (1 - \mathbb{P}_{(\Bar{\emb{\gamma}}_{o})_x}(0)).
\end{equation}
Here the notation $\mathbb{P}_{(\Bar{\emb{\gamma}}_{o})_x}(0)$ denotes the cumulative distribution function (CDF) of the rescale scalar $(\Bar{\emb{\gamma}}_{o})_x$ from $\Bar{\emb{\gamma}}_{o}$.
And thus Equation~(\ref{eq:re_l0}) provides the equivalent functionality of penalty on the number of non-zero entries that is originally posted in Formulation~(\ref{eq:inti_loss}).
Moreover, as shown in \cite{2016_concrete_distribution}, the probability distribution function (PDF) and CDF of un-rescaled scalar $({\emb{\gamma}}_{o})_x$ are respectively derivatived as:
\begin{equation}
\begin{aligned}
    p_{(\emb{\gamma}_{o})_x}(r) &= \frac{\tau (\emb{\beta}_{o})_{x} r^{-\tau-1} (1-r)^{-\tau-1}} {((\emb{\beta}_{o})_{x} r^{-\tau} + (1-r)^{-\tau})^2}, \\
      \mathbb{P}_{(\emb{\gamma}_{o})_x}(r) &= \sigma\big( (\log r - \log(1-r)) \tau - \log(\emb{\beta}_o)_x\big).
\end{aligned}
\end{equation}
Due to the monotonicity of the rescaling operation in Equation~\ref{eq:rscaling}, letting notation $g(r)$=$(1 - \eta)r + \eta$, the PDF of $(\Bar{\emb{\gamma}}_o)_x$ can be derived as:
\begin{equation}
\begin{aligned}
  p_{(\Bar{\emb{\gamma}}_{o})_x}(r) & = p_{({\emb{\gamma}}_{o})_x}(g^{-1}(r)) |\frac{\partial}{\partial r} g^{-1}(r)| \\
  & = \frac{(1 - \eta) \tau (\emb{\beta}_{o})_{x} (r-\eta)^{-\tau-1} (1-r)^{-\tau-1}} {((\emb{\beta}_{o})_{x} (r-\eta)^{-\tau} + (1-r)^{-\tau})^2}. \\
\end{aligned}
\end{equation}
Similarity, the CDF of $(\Bar{\emb{\gamma}}_{o})_x$ is:
\begin{equation}
\begin{aligned}
  \mathbb{P}_{(\Bar{\emb{\gamma}}_{o})_x}(r) & = \mathbb{P}_{({\emb{\gamma}}_{o})_x}(g^{-1}(r)) \\
& = \sigma\Big(\big(\log(r-\eta) - \log(1-r)\big) \tau - \log(\emb{\beta}_{o})_x\Big).
\end{aligned}
\end{equation}
By setting $r$ $=$ $0$, $\mathcal{L}_{SP}$ can be finally reformulated as:
\begin{equation}
\mathcal{L}_{SP} = \sum_{o \in \{\mathcal{U} \cup \mathcal{I}\}} \sum_{x=1}^{|\mathcal{S}|} \sigma\big(\log(\emb{\beta}_o)_x - \tau \log(-\eta)\big).
\end{equation}

\section{\textbf{Experiments}}
\label{exp}

We formally evaluate our proposed \model model with the aim of answering the following research questions:
\begin{itemize}[leftmargin=*]
\item \textbf{RQ1:} How does \model perform compared to the state-of-the-art counterparts in general recommendation scenarios?
\item \textbf{RQ2:} How does \model specialize in the cold-start scenario?
\item \textbf{RQ3:} How does our proposed Personalized Feature Reference Mechanism strengthen \model's performance?
\item \textbf{RQ4:} What is the effect of each model component?
\item \textbf{RQ5:} How do hyper-parameter settings affect \model?

\end{itemize}

\subsection{\textbf{Datasets}}
We include four public real-life datasets from: \textit{Last.FM} for music recommendation, \textit{Book-Crossing} for book recommendation, \textit{MIND} for news recommendation, and \textit{Dianping-Food} for restaurant recommendation. 
These four datasets are widely evaluated in recent works~\cite{2020_ckan,2021_AttentiveKG,2018_RippleNet,wu2020mind} and we report the dataset statistics in Table~\ref{tab:dataset}.
\begin{itemize}[leftmargin=*]
\item \textbf{Last.FM (Music)\footnote{\url{https://grouplens.org/datasets/hetrec-2011/}}} is provided by \textit{Last.FM} online music system which identifies music tracks as items.
The KG is public available\footnote{\url{https://github.com/hwwang55}} and built upon Microsoft Satori\footnote{\url{https://searchengineland.com/library/bing/bing-satori}}.

\item \textbf{Book-Crossing (Book)\footnote{\url{http://www2.informatik.uni-freiburg.de/~cziegler/BX/}}} is collected from a book community, which consists of trenchant book ratings.
The corresponding KG shares the similar construction process from Microsoft Satori and is free to access\footnote{\url{https://github.com/hwwang55}}.

\item \textbf{MIND (News)\footnote{\url{https://msnews.github.io./}}} is collected from anonymized behavior logs of Microsoft News website.
The construction of \textit{MIND}'s KG is based on \textit{spacy-entity-linker}\footnote{\url{https://github.com/egerber/spaCy-entity-linker}} tool and Wikidata\footnote{\url{https://www.wikidata.org/wiki/Wikidata:Main\_Page}}.

\item \textbf{Dianping-Food (Restaurant)\footnote{\url{https://www.dianping.com/}}} is provided by \textit{Dianping.com}, which including over 10 million interactions (including clicks, purchases, and adding to favorites) between around two million users and one thousand restaurants.
The corresponding KG is constructed by the internal toolkit of Meituan-Dianping Group. 
\end{itemize}

\begin{table}
\small
\caption{Statistics of datasets.}
  \label{tab:dataset}
  \centering
  \begin{tabular}{c|cccc}
    \toprule
    & Music & Book & News & Restaurant\\
    \midrule
    \midrule
    \# users & 1,872 & 17,860 & 299,955 & 2,298,698\\
    \# items & 3,846 & 14,967 & 47,034 & 1,362\\
    \# interactions & 42,346 & 139,746 & 5,090,654 & 23,416,418\\
    \midrule
    \midrule
    \# entities & 9,366 & 77,903 & 57,434 & 28,115\\
    \# relations & 60 & 25 & 62 & 7 \\
    \# KG triples & 15,518 & 151,500 & 719,776 & 160,519\\
    \bottomrule
  \end{tabular}
  \vspace{-0.2cm}
\end{table}

\begin{table*}[h]
  \caption{The average results of Top@10 recommendation task. (1) Underlines mean the second-best performance. (2) Bolds denote the empirical improvements against the second-best (underline) models. (3) $*$ indicates a statistically significant improvement over the second-best models over 95\% confidence level with the Wilcoxon signed-rank test.}
  \label{tab:topk_result}
  \centering
  \begin{tabular}{c|cc|cc|cc|cc}
    \toprule
   \multirow{2}{*}{Model} & \multicolumn{2}{c|}{Music} & \multicolumn{2}{c|}{Book} & \multicolumn{2}{c|}{News} & \multicolumn{2}{c}{Restaurant} \\ 
    & \textit{Recall@10}(\%) & \textit{P@10}(\%) & \textit{Recall@10}(\%) & \textit{P@10}(\%) & \textit{Recall@10}(\%) & \textit{P@10}(\%) & \textit{Recall@10}(\%) & \textit{P@10}(\%) \\
    \midrule
    \midrule
    BPRMF & 13.11 $\pm$ 0.26 & 3.12 $\pm$ 0.12 & 2.01 $\pm$ 0.11 & 0.48 $\pm$ 0.05 & 4.11 $\pm$ 0.13 & 0.81 $\pm$ 0.03 & 10.02 $\pm$ 0.60 & 2.44 $\pm$ 0.07 \\
    NFM & \,\,\,7.15 $\pm$ 1.83 & 1.64 $\pm$ 0.44 & 4.65 $\pm$ 2.92 & 1.00 $\pm$ 0.54 & 4.80 $\pm$ 0.12 & 0.97 $\pm$ 0.03 & 12.37 $\pm$ 1.22 & 2.82 $\pm$ 0.24 \\
    LightGCN & 18.90 $\pm$ 0.69 & 4.61 $\pm$ 0.20 & 5.75 $\pm$ 0.42 & 1.38 $\pm$ 0.04 & 4.93 $\pm$ 0.22 & 0.99 $\pm$ 0.05 & 14.56 $\pm$ 0.31 & 3.50 $\pm$ 0.07 \\
    CKE & 12.08 $\pm$ 1.27 & 2.85 $\pm$ 0.22 & 1.70 $\pm$ 0.11 & 0.42 $\pm$ 0.07 & 3.36 $\pm$ 0.50 & 0.67 $\pm$ 0.11 & 11.30 $\pm$ 0.78 & 2.77 $\pm$ 0.19 \\
    RippleNet & 11.71 $\pm$ 0.29 & 2.79 $\pm$ 0.06 & 4.98 $\pm$ 1.51 & 1.11 $\pm$ 0.29 & 4.20 $\pm$ 0.48 & 0.87 $\pm$ 0.09 & 12.58 $\pm$ 0.70 & 2.81 $\pm$ 0.12 \\
    KGAT & 10.25 $\pm$ 0.65 & 2.50 $\pm$ 0.16 & 2.82 $\pm$ 0.18 & 0.57 $\pm$ 0.04 & 2.72 $\pm$ 0.37 & 0.54 $\pm$ 0.07 & \underline{16.11} $\pm$ 0.38 & \underline{3.87} $\pm$ 0.05 \\
    CKAN & 12.66 $\pm$ 0.57 & 3.10 $\pm$ 0.09 & 3.06 $\pm$ 0.42 & 0.80 $\pm$ 0.09 & \underline{4.95} $\pm$ 0.18 & 0.98 $\pm$ 0.05 & 15.81 $\pm$ 0.23 & 3.70 $\pm$ 0.05 \\
    CG-KGR & 12.19 $\pm$ 0.84 & 2.87 $\pm$ 0.17 & 6.12 $\pm$ 0.72 & 1.33 $\pm$ 0.11 & 4.94 $\pm$ 0.35 & \underline{0.99} $\pm$ 0.07 & 15.90 $\pm$ 0.78 & 3.59 $\pm$ 0.24 \\
    MetaKG & \,\,\,6.87 $\pm$ 0.45 & 1.77 $\pm$ 0.11 & 3.17 $\pm$ 0.26 & 0.48 $\pm$ 0.03 & 2.33 $\pm$ 0.16 & 0.46 $\pm$ 0.04 & \,\,\,6.25 $\pm$ 0.48 & 1.25 $\pm$ 0.15 \\
    KGPL & \underline{19.75} $\pm$ 0.74 & \underline{4.70} $\pm$ 0.14 & \underline{6.43} $\pm$ 0.32 & \underline{1.38} $\pm$ 0.02 & 4.75 $\pm$ 0.19 & 0.95 $\pm$ 0.05 & 13.67 $\pm$ 0.44 & 2.92 $\pm$ 0.10 \\
    \midrule
    \midrule
    \model & $\textbf{20.52}^*$ $\pm$ 0.46 & $\textbf{4.96}^*$ $\pm$ 0.09 & $\textbf{7.27}^*$ $\pm$ 0.41 & $\textbf{1.58}^*$ $\pm$ 0.10 & $\textbf{4.99}$ $\pm$ 0.14 & $\textbf{1.00}$ $\pm$ 0.02 & $\textbf{18.70}^*$ $\pm$ 0.56 & $\textbf{4.03}^*$ $\pm$ 0.11 \\
    $\%$ Gain & 3.90\% & 5.53\% & 13.06\% & 14.49\% & 0.81\% & 1.01\% & 16.08\% & 4.13\% \\
   \bottomrule
  \end{tabular}
\end{table*}

 \begin{figure*}[t]
 \label{fig:topk_all}
  \centering
  \includegraphics[scale=0.18]{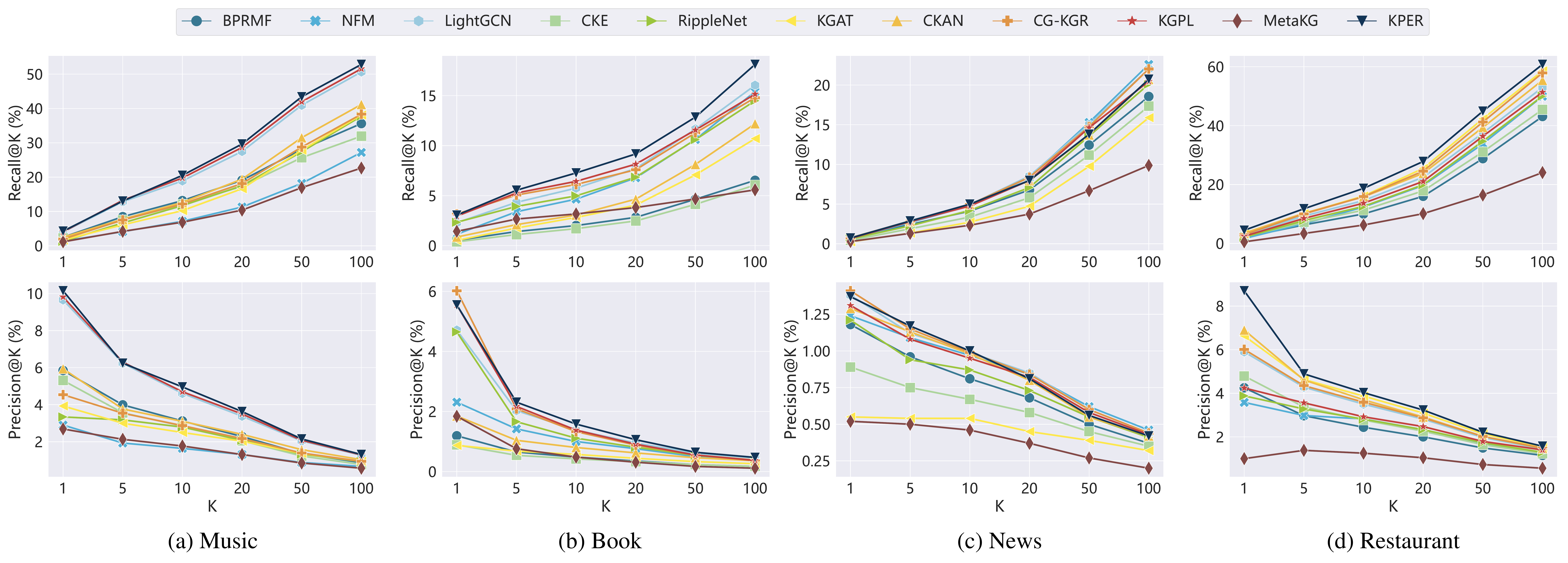}
    \caption{Average results of \textit{Recall@K} and \textit{Precision@K} in Top-K recommendation task (best view in color).}
  \label{fig:topk_main}
\end{figure*}

\subsection{\textbf{Baselines}}

We compare \model with two streams of recommender models: previous non-KG CF-based models (BPRMF, NFM, LightGCN) and recent KG-based recommender models (CKE, RippleNet, KGAT, CKAN, CG-KGR, KGPL, MetaKG).
\begin{itemize}[leftmargin=*]
    \item \textbf{BRPMF}~\cite{2009_BPRMF} is a classical CF-based method that performs matrix factorization with a pairwise ranking loss.
    
    \item \textbf{NFM}~\cite{2017_NFM} is a neural factorization machine model for collaborative filtering which utilizes the multi-layer perceptrons to model the feature interaction.
    
    \item \textbf{LightGCN}~\cite{he2020lightgcn} is a recent recommender model that achieves the state-of-the-art performance.
    It learns user-item representations with graph convolution networks and omits feature transformation and nonlinear activation, making it easier to be trained and more effective.
    
    \item \textbf{CKE}~\cite{2016_CKE} is a representative KG-based model.
    It applies TransR~\cite{2015_transR} to extract semantic representations from structural, textual and visual knowledge. 
    These semantic representations are jointly learned in a unified framework.
    
    \item \textbf{RippleNet}~\cite{2018_RippleNet} is a classical work that automatically propagates users' potential preferences along the links in KGs.
    
    \item \textbf{KGAT}~\cite{2019_KGAT} is one representative model that extends graph convolutions on KGs with an attention mechanism to jointly trained with KG embeddings with interaction embeddings.
    
    \item \textbf{CKAN}~\cite{2020_ckan} is one of the latest model that explicitly encodes collaborative signals from user-item interactions and combines them with KG associations for better recommendation.
    
    \item \textbf{CG-KGR}~\cite{2021_AttentiveKG} is another state-of-art recent method which encodes the collaborative signals as guidance in KGs for personalized recommendation.
    
    \item \textbf{MetaKG}~\cite{2022_metakg} is the other latest work that incorporates meta-learning methodology, which can be adapted to tackle the cold-start recommendation problem.
    
    \item \textbf{KGPL}~\cite{2021_KGPL} is one of the latest KG-based recommender models which employs semi-supervised learning over KGs for pseudo-labelling.
    KGRL targets for alleviating the cold-start recommendation problem.

\end{itemize} 
    

We discard potential baselines like (1) non-KG based methods, e.g., GCMC~\cite{berg2017graph}, PinSAGE~\cite{pinsage}, STAR-GCN~\cite{2019_star-gcn}, and (2) KG-based model, e.g., KGCN~\cite{2019_KGCN}, KGNN-LS~\cite{2019_KGNN_LS}, and MetaHIN~\cite{2020_metahin} methods, mainly because the previous work \cite{2020_ckan, 2022_metakg, 2021_AttentiveKG} has already validated the performance superiority.

\subsection{\textbf{Evaluation Metrics and Experiment Setup}}
We consider two evaluation tasks. (1) Top-K recommendation, and (2) Click-through rate (CTR) prediction.
\begin{itemize}[leftmargin=*]
    \item In Top-K recommendation, we apply the trained model to select K items with highest predicted probabilities for each user in the test set. 
    The performance is evaluated by \textit{Recall@K} and \textit{Precision@K}. 
    \item In CTR prediction, we use the trained model to predict the binary matching decision of each user-item pair from the test set. 
    We rescale the estimated score $\hat{y}_{u,i}$ via the sigmoid function and assign the click rate to 0 or 1 with the threshold 0.5. 
    We choose \textit{AUC} as the evaluation metric.
\end{itemize}

To present reproducible and stable results, we follow the previous work \cite{2021_AttentiveKG} to randomly split each dataset 4 times into training, validation, and test sets with the ratio of 6:2:2. 
Our proposed \model is implemented under Python 3.8 and Pytorch 1.7.1 with non-distributed training. 
The experiments are conducted on a server with Intel(R) Xeon(R) Platinum 8276, 2.20GHz CPU, 256GB RAM, and 4 NVIDIA A100 GPUs. 
For all the baselines, the hyper-parameter settings follow the original papers/codes by default or are tuned from the empirical studies. In our \model, the learning rate is tuned among \{$10^{-3}, 5\times10^{-3}, 10^{-2}, 5\times10^{-2}$\}. And we search $\eta$, $\lambda_1$, $\lambda_2$ within the range of \{-1,-0.9,-0.8,$\cdots$,-0.1\}, \{$10^{-4}, 10^{-3}, 10^{-2}$\}, and \{$10^{-5}, 10^{-4}, 10^{-3}$\}, respectively. 
We initialize and optimize all models with default Xavier initializer~\cite{2010_xavier} and Adam optimizer~\cite{2014_adam}.




\subsection{\textbf{Overall Comparison (RQ1)}}
\label{sec:rq1}
\subsubsection{Top-K Recommendation}

We evaluate \model on Top-K recommendation by varying $K$ in $\{1, 5, 10, 20, 50, 100\}$. To obtain a more detailed comparison and analysis, we first report the Top@10 recommendation results in Table \ref{tab:topk_result}, and then curve the complete Top@K results in Fig.~\ref{fig:topk_main}. 
We have the following observations:
\begin{itemize}[leftmargin=*]
    \item \textbf{Incorporating KGs is effective to improve the model performance in general Top-K recommendation.}
    The performance comparison between KG-based models and non-KG based counterparts confirms the helpfulness of incorporating KGs to compensate for semantic features in collaborative filtering. 
    (1) On one hand, for non-KG based methods, NFM and LightGCN outperform BPRMF in the most of datasets, mainly because their neural networks can effectively capture higher-order interaction information and thus facilitate the performance improvement. 
    (2) On the other hand, while several knowledge-aware methods (i.e., KGAT, CKAN and CG-KGR) perform well among all the baselines, the recent work KGPL performs the best in Music and Book datasets. 
    One explanation is that KGPL is capable of giving higher credits to the Top-K positive samples in sparse and small-scaled datasets, via their proposed pseudo-labelling approach, as we introduced in~\cref{related_work}. 
    However, for lager datasets, i.e, News and Restaurant, such exhaustive label assignments is inefficient and may be difficult to guarantee the validity of the labels.


    \

    \item \textbf{The results of Top-10 recommendation prove that \model achieves significant and stable performance improvement.}
     (1) As shown in Table \ref{tab:topk_result}, on Music, Book, News, and Restaurant datasets, \model improves over the state-of-the-art methods \textit{w.r.t.} \textit{Recall@10} by 3.90\%, 13.06\%, 0.81\%, and 16.08\%, as well as \textit{Precision@10} by 5.53\%, 14.49\%, 1.01\%, and 4.13\%. 
     The corresponding standard deviations suggest that our model's results are on the same level of stability as these baselines. 
     (2) Moreover, the Wilcoxon signed-rank tests also verify that the \model's improvements over the second-best model across most datasets are statistically significant over 95\% confidence level.
     
     \
    
    \item \textbf{As $K$ increases, \model consistently shows the competitive performance compared to the baselines.}
    (1) By performing our proposed \textit{candidate probing} in the embedding space, where both structural and semantic information are well encoded, our model \model is proved to be effectively capable of learning representative feature enrichment for to user-item representations.
    (2) Furthermore, we notice that the improvement of \model on Book and Restaurant datasets are higher than that of on Music and News datasets. 
    This demonstrates the effectiveness of our proposed model for ``cold'' user recommendation with sparse interactions, i.e., $\frac{\# \text{interactions}}{\#\text{users}}$.
    For example, users in Book and Restaurant present 7.82 and 10.19 average interactions, while for the other two dataset Music and News, the average interactions are 22.62 and 16.97.
    In~\cref{sec:cold_start_exp}, we formally analyze \model performance in cold-start recommendation scenario.
  
\end{itemize}

\begin{table}[t]
  \scriptsize
  \caption{The average results of CTR prediction task. (1) Underlines mean the second-best performance. 
  (2) Bolds denote the empirical improvements against the second-best (underline) models.
  (3) $*$ indicates a statistically significant improvement over the second-best models over 95\% confidence level with the Wilcoxon signed-rank test.}
  \setlength{\tabcolsep}{2mm}{
  \label{tab:ctr_result}
  \centering
  \begin{tabular}{c|c|c|c|c}
    \toprule
    \multirow{2}{*}{Model} & Music & Book & News & Restaurant \\ 
    & \textit{AUC}(\%) & \textit{AUC}(\%) & \textit{AUC}(\%) & \textit{AUC}(\%) \\
    \midrule
    \midrule
    BPRMF & 79.49 $\pm$ 0.44 & 65.12 $\pm$ 0.51 & 93.85 $\pm$ 0.05 & 81.81 $\pm$ 0.03 \\
    NFM & 80.09 $\pm$ 0.26 & 73.75 $\pm$ 0.61 & 94.69 $\pm$ 0.02 & 87.26 $\pm$ 0.01 \\
    CKE & 83.36 $\pm$ 0.18 & 64.40 $\pm$ 0.07 & 94.33 $\pm$ 0.01 & 83.14 $\pm$ 0.09 \\ 
    RippleNet & 81.00 $\pm$ 0.49 & 71.44 $\pm$ 1.03 & 94.68 $\pm$ 0.03 & 85.54 $\pm$ 0.17 \\
    KGAT & 83.38 $\pm$ 0.26 & 71.19 $\pm$ 0.27 & 94.78 $\pm$ 0.08 & 87.22 $\pm$ 0.15 \\
    CKAN & 83.86 $\pm$ 0.20 & 74.63 $\pm$ 0.27 & 93.25 $\pm$ 0.09 & 87.77 $\pm$ 0.01 \\
    CG-KGR & 82.52 $\pm$ 0.42 & \underline{75.73} $\pm$ 0.31 & \textbf{95.26} $\pm$ 0.02 & \textbf{89.58} $\pm$ 0.04\\
    LightGCN & 84.10 $\pm$ 0.14 & 67.02 $\pm$ 0.25 & 93.70 $\pm$ 0.01 & 82.98 $\pm$ 0.05 \\
    MetaKG & 84.24 $\pm$ 0.19 & 71.13 $\pm$ 0.43 & 93.50 $\pm$ 0.03 & 86.65 $\pm$ 0.77 \\
    KGPL & \underline{84.40} $\pm$ 0.38 & 73.64 $\pm$ 0.47 & 94.46 $\pm$ 0.09 & 84.43 $\pm$ 0.09 \\
    \midrule
    \midrule
    \model & $\textbf{87.45}^*$ $\pm$ 0.11 & $\textbf{78.94}^*$ $\pm$ 0.36 & \underline{94.82} $\pm$ 0.03 & \underline{89.03} $\pm$ 0.21 \\
    $\%$ Gain & 3.61\% & 4.24\% & -0.46\% & -0.61\% \\
   \bottomrule
 \end{tabular}}
\end{table}

\subsubsection{CTR Prediction}

In addition, we also validate the performance of our model on the CTR prediction task and present the results in Table \ref{tab:ctr_result}.
(1) Compared to these state-of-the-art baselines, \model achieves strongly competitive and stable performance. Specifically, \model improves the baselines on Music and Book datasets \textit{w.r.t.} \textit{AUC} by 3.61\% and 4.24\% with low variance, respectively. 
(2) On News and Restaurant datasets, \model achieves the \textbf{second-best} performance and slightly underperforms CG-KGR by -0.46\% and -0.61\%.
(3) Compared to Top-K recommendation task, this performance difference on CTR prediction is much smaller than that in Top-K recommendation. 
This is because our \model so far is trained for ranking tasks, i.e.,  Top-K recommendation, via encouraging model to learn the ranking, especially for Top-K recommendation.
On the contrary, The model CTR prediction capability can be generally trained towards the pairwise classification problem, by making the judgement whether a user-item interaction exists.
(4) To further improve the CTR prediction performance as future work, one promising direction would be integrating \textit{curriculum learning} paradigm to gradually improve the classification capability especially for hard negative samples. 



\subsection{\textbf{Comparison in Cold-Start Scenario (RQ2)}}
\label{sec:cold_start_exp}

\subsubsection{Analysis on Cold Users}
\begin{figure}[t]
  \centering
  \includegraphics[scale=0.2]{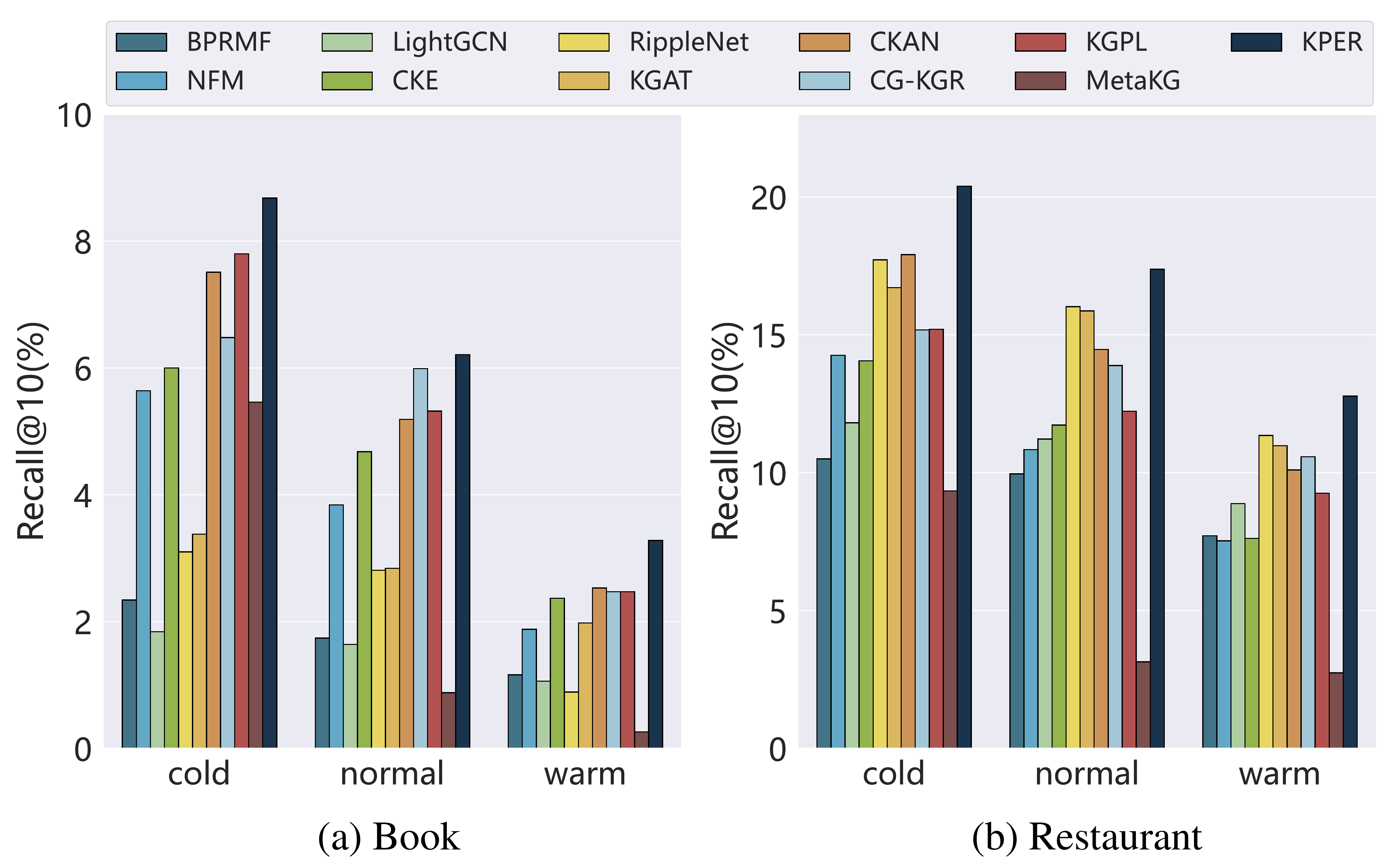}
  \caption{Average results of \textit{Recall@10} for three user groups with different sparsity level in (a) Book and (b) Restaurant. }
   \label{fig:cold_user}
\end{figure}

To investigate the performance of ``cold'' users with limited interactions, we split all users into three groups based on descending popularity, while keeping the total number of interactions for each group the same. The sparsity decreases from group-1 to group-3. We consider users in group-3 as cold users, group-2 for normal users, and group-1 for warm users.
We evaluate on the task of Top-K recommendation. Due to space limitation, we report experimental results for the two most sparse datasets (i.e., $\frac{\# \text{interactions}}{\#\text{users}}$ = 7.82 and 10.19 for Book and Restaurant) in Fig.~\ref{fig:cold_user}.

 As shown in Fig.~\ref{fig:cold_user}, (1) \model considerably improves the recommendation performance for users of all three sparsity levels, in which the cold users benefit the most.
 This verifies the efficacy of probing similar user candidates in the semantic space for enhancing cold-start recommendation performance.
 (2) Compared to other KG-aware recommender systems (e.g., KGAT, CKAN, KGPL), our \model consistently shows the performance superiority over them. 
 This shows insufficient capability of these models in cold-start scenario, indicating that a straightforward model adaptation may not be effective.
 



\subsubsection{Analysis on Cold Items}



 
We also justify the effectiveness of \model for ``cold'' items with limited popularity, we follow \cite{2016_coverage} to adopt \textit{Popularity Stratified Recall@K} (denoted as \textit{PSR@K}) metric as follows:
\begin{equation}
    \frac{\sum_{u\in\mathcal{U}} \sum_{i\in\mathcal{I}_u^{rec} \cap \mathcal{I}_u^{test}} (\frac{1}{N_i^+})^\beta}{\sum_{u\in\mathcal{U}} \sum_{i\in \mathcal{I}_u^{test}} (\frac{1}{N_i^+})^\beta} , \nonumber
\end{equation}
where $\mathcal{I}_u^{rec}$ and $\mathcal{I}_u^{test}$ are user $u$'s predicted items in the Top-K recommendation list and historical interacted items, respectively.
$N_i^+$ is the number of \textit{relevant ratings} for item $i$, and $\beta$ is a hyperparameter that adjusts for the popularity bias.
Higher values of \textit{PSR@K} indicate that more relevant cold items are recommended. 
In our experiments, we used $\beta=0.1$.
The corresponding results are reported in Table \ref{tab:cold_item}, where the Top-K \textit{PSR} values on four datasets are respectively denoted by M-\textit{PSR@10}, B-\textit{PSR@10}, N-\textit{PSR@10}, and R-\textit{PSR@10}.

\begin{table}[t]
  \caption{The PSR@10 results. Underlines and bolds denote the second-best and the best model performance.}
  \label{tab:cold_item}
    \centering
\scalebox{1.0}{
  \begin{tabular}{c|c|c|c|c}
    \toprule
   Model & M-\textit{PSR@10} & B-\textit{PSR@10} & N-\textit{PSR@10} & R-\textit{PSR@10} \\
    \midrule
    \midrule
    BPRMF & 9.57 & 1.54 & 2.90  & \,\,\,8.62 \\
    NFM & 4.99  & 3.69  & 3.20 & 10.33  \\
    LightGCN & 12.05  & 4.24 & 3.31  & 12.89  \\
    CKE & 9.85 & 1.37  & 2.41 & \,\,\,9.65  \\
    RippleNet & 7.62 & 3.81  & 2.77  & 11.36  \\
    KGAT & 7.92  & 1.98  & 2.02  & \underline{14.21}  \\
    CKAN & 9.51  & 2.46  & 3.30 & 14.19  \\
    CG-KGR & 8.65  & 4.37 & \underline{3.33} & 13.93  \\
    MetaKG & 6.02  & 3.09  & 1.74  & \,\,\,5.84  \\
    KGPL & \underline{13.43}  & \underline{4.65}  & 3.26  & 12.59 \\
    \midrule
    \midrule
    \model & \textbf{14.15} & \textbf{5.27} & \textbf{3.37} & \textbf{16.74} \\
    $\%$ Gain & 5.36\% & 13.33\% & 1.20\% & 17.80\% \\
   \bottomrule
  \end{tabular}}
\end{table}

As shown in Table \ref{tab:cold_item}, graph-based methods (e.g., LightGCN, KGAT, CG-KGR, and KGPL) generally achieve competitive performance. This is because the graph neural networks are able to capture higher-order node proximity and thus are beneficial to model implicit relations in the cold-start scenario.
But most importantly, compared to these baseline models, \model significantly improves the performance \textit{w.r.t.} \textit{PSR@10} by 5.36\%, 13.33\%, 1.20\%, 17.80\%, respectively. 
This shows that \model is capable to retrieve those \textit{temporarily less popular} \textit{actually relevant} items to users, demonstrating its superiority in alleviating the cold-start problem.

\begin{table*}[h]
  \caption{Ablation study on Personalized Feature Referencing Mechanism.}
  \label{tab:share_result}
  \centering
  \setlength{\tabcolsep}{1.8mm}{
  \begin{tabular}{l|cc|cc|cc|cc}
    \toprule
   \multirow{2}{*}{Model} & \multicolumn{2}{c|}{Music} & \multicolumn{2}{c|}{Book} & \multicolumn{2}{c|}{News} & \multicolumn{2}{c}{Restaurant} \\ 
    & \textit{Recall@10}(\%) & \textit{P@10}(\%) & \textit{Recall@10}(\%) & \textit{P@10}(\%) & \textit{Recall@10}(\%) & \textit{P@10}(\%) & \textit{Recall@10}(\%) & \textit{P@10}(\%) \\
    \midrule
    \midrule
    $\text{\model}_{\textit{ Random}}$ & 18.94 \scriptsize{(-7.70\%)} & 4.49 \scriptsize{(-9.48\%)} & 3.16 \scriptsize{(-56.53\%)} & 0.84 \scriptsize{(-46.84\%)} & 3.07 \scriptsize{(-38.48\%)} & 0.79 \scriptsize{(-21.00\%)} & 17.25 \scriptsize{(-7.75\%)} & 3.88 \scriptsize{(-3.72\%)} \\
    $\text{\model}_{\textit{ K-means}}$ & 19.66 \scriptsize{(-4.19\%)} & 4.31\scriptsize{(-13.10\%)} & 4.77 \scriptsize{(-34.39\%)} & 1.08 \scriptsize{(-31.65\%)} & 4.42 \scriptsize{(-11.42\%)} & 0.98 \,\,\scriptsize{(-2.00\%)} & $\textbf{18.86}$ \scriptsize{(+0.86\%)} & 3.89 \scriptsize{(-3.47\%)} \\
    $\text{\model}_{\textit{ All}}$ & 20.06 \scriptsize{(-2.24\%)} & 4.78 \scriptsize{(-3.63\%)} & 4.59 \scriptsize{(-36.86\%)} & 1.10 \scriptsize{(-30.38\%)} & 3.04 \scriptsize{(-39.08\%)} & 0.72 \scriptsize{(-28.00\%)} & 16.93 \scriptsize{(-9.47\%)} & 3.72 \scriptsize{(-7.69\%)} \\
    \midrule
    $\text{\model}_{\textit{ Partial-inter}}$ & 19.71 \scriptsize{(-3.95\%)} & 4.54 \scriptsize{(-8.47\%)} & 3.36 \scriptsize{(-53.78\%)} & 0.78 \scriptsize{(-50.63\%)} & 4.75 \scriptsize{(-4.81\%)} & 0.85 \scriptsize{(-15.00\%)} & 16.62\scriptsize{(-11.12\%)} & 3.53 \scriptsize{(-7.69\%)} \\
    $\text{\model}_{\textit{ Partial-KG}}$ & 18.59 \scriptsize{(-9.41\%)} & 4.22\scriptsize{(-14.92\%)} & 3.28 \scriptsize{(-54.88\%)} & 0.80 \scriptsize{(-49.38\%)} & 4.96 \scriptsize{(-0.60\%)} & 0.91 \,\,\scriptsize{(-9.00\%)} & 17.83 \scriptsize{(-4.65\%)} & 3.52\scriptsize{(-12.66\%)} \\
    \midrule
    \model & $\textbf{20.52}$ & $\textbf{4.96}$ & $\textbf{7.27}$ & $\textbf{1.58}$  & $\textbf{4.99}$ & $\textbf{1.00}$ & 18.70 & $\textbf{4.03}$ \\
   \bottomrule
  \end{tabular}}
\end{table*}

\subsection{\textbf{Personalized Feature Referencing Analysis (RQ3)}}
We formally investigate the effectiveness of our \textit{Personalized Feature Reference Mechanism}. 
We first provide a detailed ablation study and then give a case study for visualization.

\subsubsection{Ablation Study}


To differentiate the effect of our personalized seed probing, we provide the following analysis:
\begin{itemize}[leftmargin=*]
\item We first set three variants \modelv$_{\textit{Random}}$, \modelv$_{\textit{K-means}}$, and \modelv$_{\textit{All}}$.
\modelv$_{\textit{Random}}$ randomly samples the nodes from \textit{seed pool} $S$ (\cref{sec:Dynamic_shared_Embedding}).
\modelv$_{\textit{K-means}}$ employs K-means as the seed probing method.
\modelv$_{\textit{All}}$ brutally takes all nodes in $S$ for computation.
We fix $|S|=128$ for fair comparison.
From the results in Table~\ref{tab:share_result}, we observe that \modelv$_{\textit{K-means}}$ performs better than \modelv$_{\textit{Random}}$ in most cases.
This generally follows the intuition that centroids-nearby nodes are usually more representative to present unique semantics than other randomly selected ones.
Furthermore, although \modelv$_{\textit{All}}$ performs slightly better than \modelv$_{\textit{Random}}$ on Music and Book datasets, it underperforms others on larger datasets, i.e., News and Restaurant, while not only may introduce noise but also increase the heavy computational burden to the model training. 
By contrast, our proposed adaptive approach makes a good balance between \textit{performance stability} across different datasets and \textit{seed allocation concision} for computation ease.

\vspace{0.4em}

\item We also study how the encoded information affects the seed probing process. 
We introduce two variants, i.e., \modelv$_{\textit{Partial-inter}}$ and \modelv$_{\textit{Partial-KG}}$, by partially using one side of information, i.e., from interaction side only and KG side only.
As we can observe from Table~\ref{tab:share_result}, partially utilizing information may lead to the sup-optimal model learning and thus impede \model performance. 
\end{itemize}


\subsubsection{Case Study}
\begin{figure}[htbp]
  \centering
  \includegraphics[scale=0.32]{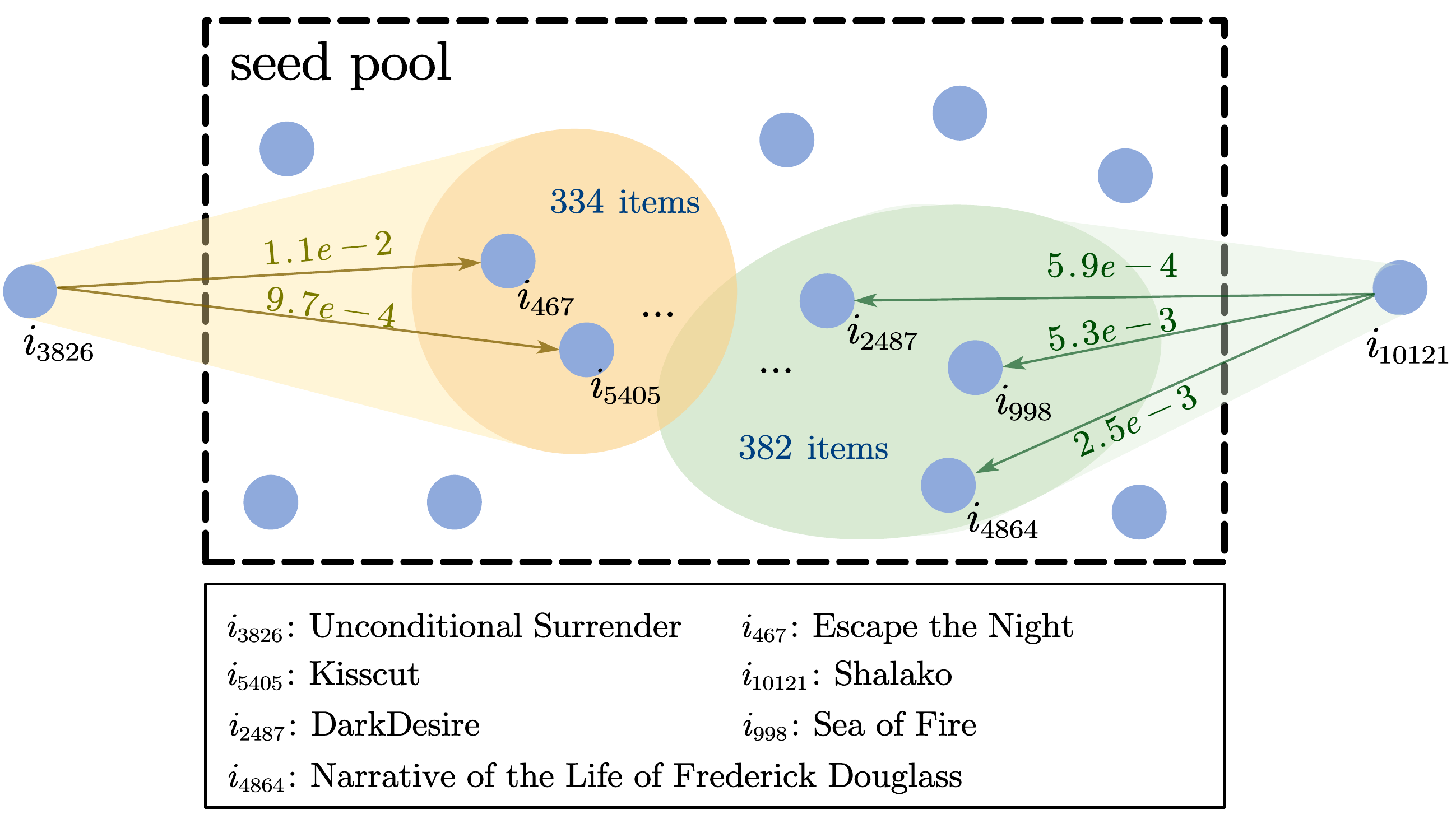}
    \caption{A real example from Book dataset.}
   \label{fig:casestudy}
\end{figure}

To visualize the effect of our proposed \textit{Adaptive Seed Probing}, we take a real case from Book dataset in Fig.~\ref{fig:casestudy}. 
By adaptive seed probing, our model can well customize to find 334 and 382 candidates for item $i_{3826}$ and item $i_{10121}$, respectively. 
By attentively assigning the weight 0.011 item $i_{3826}$ to seed $i_{467}$, \model highlights the strong confidence of information referencing and pays a less attention to seed $i_{5405}$ with a lower weight 0.00097. 
This example visually shows that, endorsed by our proposed Adaptive Seed Probing approach, \model can well distinguish the semantically similar nodes (i.e., seed $i_{467}$) out of dissimilar one (i.e., seed $i_{3826}$). 
As for the item $i_{10121}$, the weights regarding to three seeds (i.e., $i_{2487}$, $i_{998}$, and $i_{4864}$) show the similar confidences and thus these three seeds equally contribute to information reference for the item $i_{10121}$.

\subsection{\textbf{ Ablation Study of \model (RQ4)}}
To validate the effectiveness of different parts in \model, we conduct a comprehensive ablation study on the task of Top-K recommendation task and report the results in terms of \textit{Recall@10} and \textit{Precision@10} on four datasets.
\begin{table}[h]
  \scriptsize
  \caption{Ablation study on Top@10 Recommendation.}
  \label{tab:abl_result}
  \centering
  \setlength{\tabcolsep}{2.6mm}{
  \begin{tabular}{c|c|c|c|c}
    \toprule
    Dataset & w/o Inter & w/o KG-att & w/o Ref & \model \\
    \midrule
    \midrule
    M-\textit{R@10}(\%) & 18.64 \tiny{(-9.16\%)} & 17.49 \tiny{(-14.77\%)} & 19.10 \tiny{(-6.92\%)} & \textbf{20.52} \\
    M-\textit{P@10}(\%) & 4.35 \tiny{(-12.30\%)} & 4.20 \tiny{(-15.66\%)} & 4.34 \tiny{(-12.50\%)} & \textbf{4.96} \\
    \midrule
    B-\textit{R@10}(\%) & 7.16 \tiny{(-1.51\%)} & 6.49 \tiny{(-10.73\%)} & 2.90 \tiny{(-60.11\%)} & \textbf{7.27} \\
    B-\textit{P@10}(\%) & 1.45 \tiny{(-8.23\%)} & 1.27 \tiny{(-19.62\%)} & 0.72 \tiny{(-54.43\%)} & \textbf{1.58} \\
    \midrule
    N-\textit{R@10}(\%) & 4.47 \tiny{(-10.42\%)} & 4.68 \tiny{(-6.21\%)} & 3.20 \tiny{(-35.87\%)} & \textbf{4.99} \\
    N-\textit{P@10}(\%) & 0.95 \tiny{(-5.00\%)} & 0.96 \tiny{(-4.00\%)} & 0.78 \tiny{(-22.00\%)} & \textbf{1.00} \\
    \midrule
    R-\textit{R@10}(\%) & 13.15 \tiny{(-29.68\%)} & 16.02 \tiny{(-14.33\%)} & 15.47 \tiny{(-17.27\%)} & \textbf{18.70} \\
    R-\textit{P@10}(\%) & 3.63 \tiny{(-9.93\%)} & 3.58 \tiny{(-11.17\%)} & 3.60 \tiny{(-10.67\%)} & \textbf{4.03} \\
   \bottomrule
  \end{tabular}}
\end{table}

\noindent\textbf{Effect of Interactive Information Encoding.}
We first verify the effectiveness of \textit{interactive information encoding} (\cref{sec:User_Embedding}). The variant w/o Inter, discards our attentive approach to simply average the historical interacted embeddings in aggregation. 
As shown in Table \ref{tab:abl_result}, the performance degradation of w/o Inter justifies the effectiveness of interactive information in improving \model performance across all datasets. 

\noindent\textbf{Effect of Attentive Knowledge Encoding.}
We move on to studying the effect of our \textit{knowledge-aware attentive encoding} module (\cref{sec:Knowledge-aware_Attentive_Embedding}) by replacing the original attention mechanism to force the knowledge triples to contribute equally. 
We denote it as w/o KG-att. As we can observe, variant w/o KG-att confronts a conspicuous performance decay, which shows that our knowledge-aware attentive encoding provides an effective approach to extract knowledge from KGs and thus benefits the recommendation performance.

\noindent\textbf{Effect of Personalized Feature Referencing.}
We use a variant, i.e., w/o Ref, To substantiate the impact of our \textit{Personalized Feature Referencing} (\cref{sec:Dynamic_shared_Embedding}), by totally removing this module in model training. 
From Table \ref{tab:abl_result}, we find that the performance of w/o Ref dramatically degrades across all datasets.
This supports that our proposed feature referencing is a powerful determinant, enabling \model to detect potential similar candidates for each user/item, and further improve the recommendation performance.

\subsection{\textbf{Hyper-parameter Analysis (RQ5)}}
\noindent\textbf{Sampling in Interactive Information Encoding.}
We investigate the effect of neighbor sampling in interactive information encoding, and take user nodes are examples to conduct the experiments. 
The results are presented in the left-part of Table~\ref{tab:hyper_sample_result}. 
We notice that: 
a larger number of sample size, denoted as $l$, (or even the whole node neighborhood) is not a guarantee for a better performance. The best performance is obtained when $l$ is 16, 8, 8, 16, respectively. 
One reasonable explanation is that heavy repetitive sampling on sparse datasets leads to overfitting, which has the negative impact on the model training.

\begin{table}[htbp]
  \caption{Varying sample size $l$ (left) and knowledge extraction depth $K$ (right) in Top@10 Recommendation.}
  \label{tab:hyper_sample_result}
  \centering
\setlength{\tabcolsep}{1mm}{
  \begin{tabular}{c|c|c|c}
    \toprule
    Dataset & $l=8$ & $l=16$ & $l=32$ \\
    \midrule
    \midrule
    M-\textit{R@10}(\%) & 19.58 & \textbf{20.52} & 19.40 \\
    M-\textit{P@10}(\%) & 4.25 & \textbf{4.96} & 4.51 \\
    \midrule
    B-\textit{R@10}(\%) & \textbf{7.27} & 6.14 & 7.19 \\
    B-\textit{P@10}(\%) & \textbf{1.58} & 1.12 & 1.42 \\
    \midrule
    N-\textit{R@10}(\%) & \textbf{4.99} & 4.86 & 4.14 \\
    N-\textit{P@10}(\%) & \textbf{1.00} & 0.96 & 0.71 \\
    \midrule
    R-\textit{R@10}(\%) & 18.11 & \textbf{18.70} & 17.36 \\
    R-\textit{P@10}(\%) & 3.89 & \textbf{4.03} & 3.79 \\
   \bottomrule
  \end{tabular}}
    \setlength{\tabcolsep}{1mm}{
    \begin{tabular}{c|c|c|c}
    \toprule
    $K=0$ & $K=1$ & $K=2$ & $K=3$ \\
    \midrule
    \midrule
    16.36 & 19.41 & \textbf{20.52} & 18.93 \\
    3.94 & 4.25 & \textbf{4.96} & 4.05 \\
    \midrule
    4.05 & 6.49 & \textbf{7.27} & 5.93  \\
    1.10 & 1.41 & \textbf{1.58} & 1.30 \\
    \midrule
    4.24 & \textbf{4.99} & 4.64 & 4.68 \\
    0.73 & \textbf{1.00} & 0.96 & 0.85 \\
    \midrule
    15.38 & \textbf{18.70} & 18.46 & 18.25  \\
    3.14 & \textbf{4.03} & 3.90 & 4.01  \\
   \bottomrule
  \end{tabular}}
  \vspace{-0.1cm}
\end{table}

\noindent\textbf{Depth of Attentive Knowledge Encoding.}
 In the second part of Table \ref{tab:hyper_sample_result}, we further present the evaluation results of different knowledge extraction depths, by varying $K$ from 0 to 3 and 0 means no information aggregated from KGs. 
 \model achieves the best performance when $K$ is 2, 2, 1, and 1 for all benchmarks, respectively. 
 One possible reason for this phenomenon is that a longer distance information propagation may incur more noise, especially on large datasets and maintaining a appropriate depth of knowledge encoding can ultimately maximize the recommendation performance.


\section{\textbf{Conclusion \& Future Work}}
\label{conclusion}

In this paper, we propose a novel knowledge-graph-based recommender model \model, towards the alleviation of cold-start problem.
The core mechanism, namely \textit{Personalized Feature Referencing}, enables an effective latent feature enrichment to user-item representations, while getting rid of the interaction sparsity constraint.
The extensive experiments well demonstrate the model effectiveness in both general and cold-start recommendation scenarios.

In the future, we plan to investigate two major possible directions. 
(1) It is worth further improving \model by adopting different learning paradigms, e.g., \textit{curriculum learning}.
(2) We may extend our methodology to session recommendation\cite{2022_DAGNN} or sequential recommendation\cite{chen2020sequence}, in which modeling the sequential interactive trajectory may be harder to capture and predict in cold-start scenarios.

\ifCLASSOPTIONcaptionsoff
  \newpage
\fi
\bibliographystyle{IEEEtran}
\bibliography{ref}


\end{document}